\documentclass[aps,manuscript]{revtex4} \usepackage{graphics}
\input epsf
\def\ba{\begin{eqnarray}}
\def\ea{\end{eqnarray}}
\def\be{\begin{equation}}
\def\ee{\end{equation}}

\def\mxth{\mathsurround=0pt }
\def\xversim#1#2{\lower2.pt\vbox{\baselineskip0pt \lineskip-.2pt
    \ialign{$\mxth#1\hfil##\hfil$\crcr#2\crcr\sim\crcr}}}

\newcommand{\labeq}[1] {\label{eq:#1}}

\newcommand{\labfig}[1] {\label{fig:#1}}  

\begin{document}

\title{M Theory Model of a Big Crunch/Big Bang Transition}

\author{
Neil Turok$^1$, Malcolm Perry$^1$, and Paul J. Steinhardt$^2$}

\affiliation{~}

\affiliation{$^1$DAMTP, Centre for Mathematical Sciences,
Wilberforce Road, Cambridge CB3 0WA, UK }
\affiliation{$^2$Joseph Henry Laboratories,
Princeton University,
Princeton, NJ 08544, USA,\\ 
Institute for Advanced Studies, Olden Lane, Princeton, NJ 08540, USA.}

\begin{abstract}
We consider a picture in which the transition from a big crunch to a big bang
corresponds to the collision of two empty orbifold planes approaching each
other at a constant non-relativistic speed in a locally flat
background space-time, a situation relevant to 
recently proposed cosmological
models. We show that $p$-brane states which wind around the extra dimension
propagate smoothly and unambiguously across the orbifold plane collision.
In particular we calculate the quantum mechanical production of winding 
M2-branes extending from one orbifold to the other.  We find
that the resulting density is finite and that the resulting
gravitational back-reaction is
small.  These winding states, which include the string theory
graviton, can be propagated
smoothly across the transition using a perturbative expansion 
in the membrane 
tension, an expansion which from the point of view of string theory
is an expansion in 
{\it inverse} powers of $\alpha'$.  The conventional 
description of a crunch based on Einstein general relativity,
involving Kasner or mixmaster behavior is misleading, we
argue, because general relativity is only the leading order
approximation to string theory in an expansion in positive
powers of $\alpha'$. In contrast, in the M theory setup we
argue that interactions should be well-behaved because of
the smooth evolution of the fields combined with the fact
that the string coupling tends to zero at the crunch.
The production of massive Kaluza-Klein states should also be
exponentially
suppressed for small collision speeds.  
We contrast this good behavior with that found in
previous studies of strings in Lorentzian orbifolds.
\hfill\break
PACS number(s):
11.25.-w,04.50.+h, 98.80.Cq,98.80.-k
\end{abstract}

\maketitle

%
%

\renewcommand\baselinestretch{1.0}
\tableofcontents

\renewcommand\baselinestretch{1.0}

\section{Introduction}

One of the greatest challenges faced by string and M theory is
that of describing time-dependent singularities, such as 
occur in cosmology and in black holes. These singularities 
signal the catastrophic failure of general relativity at short
distances, precisely the pathology that string theory 
is supposed to cure. Indeed string theory does succeed in
removing the divergences present in perturbative quantum gravity
about flat spacetime. String theory is also known to tolerate
singularities in certain static backgrounds 
such as orbifolds
and conifolds.
However, studies within string theory thus far 
have  been unable to
shed much light on the far more interesting question of 
the physical resolution of time-dependent singularities.

In this paper we discuss M theory in one of the
simplest possible time-dependent backgrounds~\cite{kosst,steif}, 
a direct product of $d-1$-dimensional 
flat Euclidean space $R^{d-1}$ with two dimensional 
compactified Milne space-time, ${\cal M}_C$,
with
line element
\be
-dt^2+t^2 d\theta^2.
\labeq{line}
\ee
The compactified coordinate $\theta$ 
runs from $0$ to $\theta_0$. 
As $t$ runs from $-\infty$ to $+\infty$, the 
compact dimension shrinks away and reappears once
more, with 
rapidity $\theta_0$. 
Analyticity in $t$ suggests that this continuation
is unique~\cite{tolley1}.

Away from $t=0$, 
${\cal M}_C$ is locally flat, as can be 
seen by changing to coordinates $T=t\cosh \theta$, $Y=t\sinh \theta$
in which (\ref{eq:line}) is just $-dT^2+dY^2$. Hence
${\cal M}_C\times R^{d-1}$ is naturally a 
solution of any geometrical theory
whose field equations are 
built from 
the curvature tensor.
However, ${\cal M}_C\times R^{d-1}$ is nonetheless
mathematically singular at $t=0$ 
because the metric 
degenerates when the compact dimension disappears.
General relativity cannot
make sense of this situation since
there ceases to be enough Cauchy data to
determine the future evolution
of fields. In fact, the situation is worse than this:
within general relativity,
generic perturbations diverge as log$|t|$ as
one approaches the singularity~\cite{ekperts},
signaling the breakdown
of perturbation theory and the approach to
Kasner or mixmaster behavior, according to which the
space-time
curvature diverges as $t^{-2}$. 
Of course, this breakdown of general
relativity presents a challenge: 
can M theory make sense of the singularity at $t=0$?

We are interested in 
what happens in the immediate
vicinity of $t=0$, when the compact dimension 
approaches, and becomes smaller than,
the fundamental membrane tension scale.
The key
difference between M theory (or string theory) and local field theories
such as general relativity is the existence of extended
objects including those stretching across
compactified  dimensions.
Such states become very light as the compact
dimensions shrink below the fundamental scale. 
They are known to play a central
role in resolving
singularities for example in orbifolds and in 
topology-changing transitions~\cite{greene}. Therefore, it
is very natural to ask what role such states
play in big crunch/big bang space-times.

In this paper we shall show 
that 
$p$-branes winding uniformly around the compact 
dimension 
obey equations, obtained by canonical methods,
which are completely
regular at $t=0$. These methods are naturally
invariant under choices of worldvolume coordinates.
Therefore we claim that it is possible to 
unambiguously describe evolution of such states
from $t<0$ to $t>0$, through a cosmological
singularity from the point of view of the low energy
effective theory.
Indeed the space-time
we consider corresponds locally to one where
two empty, flat, parallel orbifold planes collide, precisely
the situation envisaged in recently proposed cosmological
models.

Hence, the calculations we report are directly relevant to the 
ekpyrotic~\cite{kost} and
cyclic universe~\cite{STu} scenarios, in which passage
through a singularity of this type is taken 
to represent the standard  hot big bang.
In particular, the equation of state during the dark energy and
contracting phases causes the orbifold planes to be empty,
flat, and parallel as they approach to within a string 
length~\cite{chaos,review}.  
This setup makes it natural to split the study
of the collision of orbifold planes into a separate  analysis of the
winding modes, which become light near $t=0$, and 
other modes that become heavy there.  This strategy feeds directly into
the considerations in this paper.

The physical reason why winding states are well-behaved is easy to
understand. The obvious problem with a space-time such as
compactified Milne is the blue shifting effect felt by 
particles which can 
run around the compact dimension as it shrinks away.
As we shall discuss in detail, winding
states wrapping around the compact dimension 
do not feel any blue shifting effect because there
is no physical motion 
along their length. Instead, as their length disappears,
from the point of view of the noncompact dimensions, 
their effective mass or tension 
tends to zero but their energy and momentum remain finite. When 
such states are quantized the corresponding fields are well-behaved
and the field equations are analytic at $t=0$.
In contrast, for bulk, non-winding states, the 
motion in the $\theta$ direction is physical and it becomes
singular as $t$ tends
to zero. In the quantum field theory of such states,
this behavior results in 
logarithmic divergences of the
fields near $t=0$, even for the lowest 
modes of the field which
are uniform in $\theta$ {\it i.e.,}  the lowest Kaluza-Klein 
modes (see Section X and Appendix 4).

We are specially interested in the case of M theory,
considered as the theory of branes. As
the compact dimension becomes small, the
winding M2-branes we focus on 
are the lowest energy states of the theory, and describe
a string theory in a certain time-dependent background. 
The most remarkable feature of this setup is that the
string theory includes a graviton and, hence, 
describes perturbative gravity near $t=0$.
In this paper, we show these strings,
when considered as winding M2-branes,
follow smooth evolution (see Section VII) across the singularity,
even though the string frame metric degenerates there. 
Furthermore we show that this good behavior is only
seen in a perturbation expansion in the membrane tension,
corresponding from the string theory point of view to an expansion
in {\it inverse powers of $\alpha'$}. We argue that
the two-dimensional 
nonlinear sigma model describing this situation
is renormalizable in such an expansion.
The good behavior of the relevant string theory
contrasts sharply with the bad behavior of general
relativity. There is no contradiction, however,
because general relativity is only the first approximation
to string theory in an expansion in positive
powers of $\alpha'$. Such an expansion is valid 
when $t$ is much larger than the fundamental membrane scale,
but it fails near the singularity where,  as mentioned,
the theory is regular in the opposite $(\alpha')^{-1}$ expansion.
The logarithmic divergences of perturbations found 
using the Einstein equations
are, thereby,  seen  to be due to the failure
of the $\alpha'$ expansion, and not of M or string theory {\it per se}.

When the M theory dimension is small, 
the modes of the theory are neatly partitioned
into light $\theta$-independent modes and heavy
$\theta$-dependent modes. 
The former set consists of winding membranes, which
describe a string theory including perturbative gravity.
This is the sector
within which cosmological perturbations
lie, and which will be our prime focus in this paper. 

The $\theta$-dependent modes are likely to be harder to 
describe. 
The naive argument that these modes are problematic
because they are  blue shifted and, hence, infinitely 
amplified as $t\rightarrow 0$ 
is suspect because it
relies on conventional
Einstein gravity,
Here we argue that,
close to the brane collision, Einstein gravity 
is a poor approximation and,  instead, perturbative
gravity is described by
the non-singular winding sector.  The latter
does not 
exhibit blue shifting behavior near $t=0$, 
so the naive argument does not apply.

Witten
has argued \cite{witten}  that 
the massive Kaluza-Klein modes of the
eleven-dimensional
theory
map onto non-perturbative black hole states in the effective
string theory.
Even though these black hole states are likely to be
hard to describe in detail, we will explain in Section II
why their overall effect is likely to be small.
First, in the cosmological scenarios of interest,
the universe enters the regime where perturbative gravity is
described by the winding modes ({\it i.e.,} the branes are close)
with a negligible density 
of Kaluza-Klein massive modes.  This suppression is a result of 
the special
equation of state in the contracting phase that precedes this
regime~\cite{review}.  Second, the density of black holes quantum produced
due to the time-dependent background in the
vicinity of the collision is likely to be 
negligible because they are so massive
and so large.

For these reasons, we focus at present on the propagation
of the perturbative gravity sector near $t=0$
corresponding to the winding M2-brane states.
In Section II, we introduce the compactified Milne background
metric that describes the collision between orbifold planes in the 
big crunch/big bang transition.  We also discuss the motivation for
the initial conditions that will be assumed in this paper.
The canonical Hamiltonian description of $p$-branes in
curved space is given in Section III and applied to winding modes
in the compactified Milne background in Section IV.  
Section V discusses the key difference in the Hamiltonian 
description between winding and bulk states that accounts for
their different behavior near the big crunch/big bang 
transition.  

Section VI is the consideration of a toy model in which 
winding strings are produced as the
branes collide.  The winding modes
are described semi-classically, and their quantum 
production at the bounce is computed.  Section VII presents the
analogous semi-classical description of winding M2-branes.  
Although we cannot solve the  theory 
exactly, we show the eleven-dimensional theory
is well behaved near $t=0$
and explain how the apparent singularity in the dimensionally-reduced
string theory is resolved in the membrane picture.  Then, Section VIII 
makes clear the difference between our calculation, an expansion in inverse
powers of $\alpha'$, versus Einstein gravity, the leading term in 
an expansion 
in positive powers of 
$\alpha'$.  This argument is key to explaining why we think the
transition is calculable even though it appears to be poorly 
behaved when described by Einstein gravity.  
 Section IX, then, 
uses 
Euclidean instanton methods to study 
the quantum production of
winding M2-branes (in analogy to the case of  winding strings in Section VI) 
induced by
passage through the singularity, obtaining
finite and physically sensible results. In particular,
the resultant density
tends to zero as the speed
of contraction of the compact dimension is reduced.
We  estimate the gravitational back-reaction
and show it is small provided $\theta_0$,
the rapidity of contraction of the compact dimension,
is small.

In Section X, we comment on why our M theory setup  
is better behaved than the Lorentzian orbifold case~\cite{seibergetal}
considered in some previous investigations of the big crunch/big
bang transition. The fundamental problem with the latter case,
we argue, is that perturbative gravity lies within the
bulk sector and not the winding sector as far as
the compactified Milne singularity is concerned.
Therefore, it is susceptible to the blueshifting
problem mentioned above, rendering the string equations singular.

\section{The Background Big Crunch/Big Bang Space-time}

The $d+1$-dimensional 
space-time we consider is a direct product of 
$d-1$-dimensional Euclidean space, $R^{d-1}$, and a 
two-dimensional time-dependent space-time
known as compactified
Milne space-time, or ${\cal M}_C$. 
The line element for ${\cal M}_C \times R^{d-1}$ is thus
\be
ds^2=  -dt^2 +t^2 d\theta^2 + d\vec{x}\,^2, \qquad 0\leq \theta \leq \theta_0,
\qquad -\infty <t <\infty,
\labeq{backg}
\ee
where $\vec{x}$ are Euclidean 
coordinates on $R^{d-1}$, $\theta$ 
parameterizes the compact dimension and $t$ is the time.
The compact dimension may 
either be a circle, in which case we identify $\theta$
with $\theta+\theta_0$, or a $Z_2$ orbifold in which case we
identify $\theta$ with  $\theta+2 \theta_0$ and 
further identify $\theta$ with $2\theta_0-\theta$.
The fixed points $\theta=0$ and $\theta=\theta_0$ are
then interpreted as tensionless $Z_2$-branes approaching
at rapidity $\theta_0$, colliding at $t=0$ to re-emerge
 with the same relative rapidity.

The orbifold reduction is the case of prime interest in the
ekpyrotic/cyclic models, originally motivated by the
construction of heterotic M theory
from eleven dimensional supergravity~\cite{hw,ovrut}. 
In these models, the boundary branes possess nonzero tension.
However, the tension is a subdominant effect near $t=0$ and 
the brane collision is locally well-modeled by ${\cal M}_C \times R^{d-1}$
(See Ref.~\cite{tolleyperts}).

The line element (\ref{eq:backg}) is of particular interest
because it is locally flat and, hence, an exact solution not only of
 $d+1$-dimensional Einstein gravity but of 
any higher dimensional gravity theory whose field equations
are constructed from curvature invariants with no cosmological 
constant.
And even if a small cosmological constant
were present, it would not have a large effect locally 
so that
solutions with a similar local structure
in the vicinity of the singularity
would be expected to exist.

Consider the description of (\ref{eq:backg}) within
$d+1$-dimensional general relativity.
When the compact dimension is small, $\theta$-dependent 
states become massive and it makes sense to describe the
system using a low energy effective field theory. This may be 
obtained by the well known procedure of 
dimensional reduction.
The $d+1$-dimensional line element (\ref{eq:backg}) may be rewritten
in terms of a $d$-dimensional Einstein frame metric, $g^{(d)}_{\mu \nu}$,
and a scalar field $\phi$:
\be
ds^2 = e^{2\phi \sqrt{(d-2)/(d-1)}} d\theta^2 +
e^{-2\phi /\sqrt{(d-2)(d-1)}}g^{(d)}_{\mu \nu} d x^{\mu} d x^{\nu}.
\labeq{metr}
\end{equation}
The numerical coefficients are chosen so that
if one substitutes 
this metric into the $d+1$-dimensional Einstein 
action and assumes that $\phi$ and $g^{(d)}_{\mu \nu}$ are both
$\theta$-independent, one obtains 
$d$-dimensional Einstein gravity with 
a canonically normalized 
massless, minimally coupled scalar field $\phi$.
(Here we choose units in which the
coefficient of the Ricci scalar in the $d$-dimensional
Einstein action is ${1\over 2}$. We have also ignored Kaluza-Klein
vectors, which play no role in this argument and
are in any case projected out in the orbifold reduction.)

From the viewpoint of the low energy effective theory, the 
$d+1$-dimensional space-time ${\cal M}_C \times R^{d-1}$
 is reinterpreted as a
$d$-dimensional cosmological solution  where $t$ plays the 
role of the conformal time.
Comparing (\ref{eq:backg}) and (\ref{eq:metr}),
the $d$-dimensional  Einstein-frame metric
$g^{(d)}_{\mu \nu}=a^2\, \eta_{\mu \nu}$ with $a
\propto |t|^{1/(d-2)}$ and the scalar field  $\phi =
\sqrt{ (d-1)/(d-2)} {\rm ln}|t|$. From this point of view
$t=0$ is a space-like curvature
singularity of the standard big bang type where
the scalar
field diverges, and 
passing through 
$t=0$ would seem to be impossible.
However, by lifting to the higher dimensional viewpoint 
one sees that the
situation is not really so bad. The line element (\ref{eq:backg})
is in fact static at all times in the noncompact
directions $\vec{x}$. So for example, matter 
localized on the branes would see no blue shifting effect 
as the singularity approaches\cite{STu}. As we discuss
in detail in Section IV,
winding states do not see a blue shifting effect either.

In this paper, we consider  an M theory picture
with  two empty, flat, parallel colliding orbifold planes 
and we are interested in the dynamics of the collision region
from the point where the planes are roughly a string length apart.
The  assumed initial conditions are important 
for two reasons. First, they correspond to the simple compactified 
Milne background discussed above.
Second, as mentioned in the introduction,
this initial condition means that the
excitations neatly divide into 
light winding modes that are becoming massless
and heavy Kaluza-Klein modes that are becoming massive and decoupling
from the low-energy effective theory.  

What we want to show now is that initial conditions with negligible
heavy Kaluza-Klein modes present are naturally produced in cosmological
scenarios such a the cyclic model~\cite{STu,review}.
This justifies our focus 
on the winding modes throughout the remainder of the paper.  
However, the argument is inessential to the rest of the paper
and readers willing to 
accept the initial conditions without justification  may wish 
to 
proceed straight away to the next Section.

The cyclic model assumes a non-perturbative 
potential hat produces an attractive force between the orbifold 
planes.  When the branes are far apart, perhaps $10^4$ Planck lengths, 
the potential energy is positive and small, acting as the  dark energy
that causes the currently observed accelerated expansion.  
In the dark energy dominated
phase, the branes  stretch by a factor of two in linear dimensions 
every 14 billion years or so, causing the branes to become flat, parallel
and empty.  In the low-energy effective theory,
the total energy 
is dominated by the scalar field $\phi$ whose value determines the
distance between branes. 
As the planes draw together, the 
potential energy $V(\phi)$ of this field 
decreases and becomes increasingly negative until 
the expansion stops and a contracting phase begins.

A key point is that this contracting phase is described by
an attractor solution, which has an equation of state parameter
$w\equiv P/\rho \gg 1$. The energy density of the scalar field
$\phi$ scales as
\be
\rho_\phi \propto a^{-(d-1)(1+w)}
\labeq{rhophi}
\ee
in this phase. This is a very rapid increase, causing the density
in $\phi$
to come to dominate over curvature,
anisotropy, matter, or radiation\cite{chaos}. 
We now show that $\phi$ comes to dominate over 
the massive Kaluza-Klein modes. The latter
scale as
\be
\rho_{KK} \propto a^{-(d-1)} L^{-(d-1)/(d-2)}
\labeq{rhoKK}
\ee
where $L$ is the size of the extra dimension. The first factor
is the familiar inverse volume scaling which all particles suffer.
The second factor indicates the effective mass of the
Kaluza-Klein modes. The $d+1$-dimensional mass is 
$L^{-1}$, but this must be converted to a $d-$dimensional 
mass using the ratios of square roots of 
the $00$ components of the $d+1$-dimensional
metric and the $d$-dimensional metric. This correction 
produces the second factor in (\ref{eq:rhoKK}).

From (\ref{eq:metr}),
we have $L\propto e^{\phi\sqrt{(d-2)/(d-1)}}$. Now the key
point is that this scales much more slowly with $a$ than
the potential $V(\phi)$ which scales as $\rho_\phi$ in
(\ref{eq:rhophi}). Neglecting the scaling with $L$, 
the density of massive Kaluza-Klein modes scales as
as $\rho_\phi^{1/(1+w)}$. The final suppression of the
density of massive modes relative to the density in $\phi$ is
therefore $\sim (V_{i}/V_{f})^{w/(1+w)}$ where $V_{i}$ and $V_{f}$ are the
magnitudes of the scalar potential when the $w\gg 1$ phase
begins and ends. For large $w$, which we need in order to
obtain scale-invariant perturbations, this is 
an exponentially
large factor~\cite{STu,review}.

The massive Kaluza-Klein modes are, hence, exponentially
diluted when the $w\gg 1$ phase ends and the Milne
phase begins. During the Milne phase, 
the scalar field is massless and has an equation of state
$w=1$, so $\rho_\phi$ scales as $a^{-2(d-1)}$ as
the distance between the orbifold planes
shrinks to zero.
In this regime, the Kaluza-Klein massive mode density
scales in precisely the same way. Therefore, their
density remains an exponentially small fraction of
the total density right up to collision: meaning from the
string theory point of view that the black
hole states remain exponentially rare.
  
So we need only worry about
black holes  produced in the vicinity of the 
brane collision itself. From the point of view
of the higher dimensional theory, the oscillation
frequency of the masssive Kaluza-Klein modes 
$\omega \sim |\theta_0t|^{-1}$ 
changes adiabatically, 
$\dot{\omega}/\omega^2 \sim \theta_0 \ll 1$ for
small $\theta_0$, all the way to $t=0$. Therefore,
one expects little particle production before or
after $t=0$.
From the dimensionally reduced point of view, the
mass of the string theory black holes 
is larger than the Hubble constant $\sim t^{-1}$,
by the same factor $\theta_0^{-1}$.
From either analysis, production 
of such states should
be suppressed by 
a factor $e^{-1/\theta_0}$, making it negligible
for small $\theta_0$.

In sum, for the cosmological models of interest,
the Kaluza-Klein modes are 
 exponentially rare
when the Milne phase begins, and, since their mass increases
as the collision approaches, they should not be generated by the
orbifold plane motion.
(They effectively decouple from the low energy effective
theory.)
Hence, all the properties we want at the outset of our calculation here
are naturally achieved by the contracting phase with $w\gg1$, as occurs
in some current cosmological models\cite{review}.

\section{General Hamiltonian for $p$-Branes in Curved Space}

The classical and quantum dynamics of $p$-branes may be
treated using canonical methods, indeed $p$-branes 
provide an application {\it par
excellence} of Dirac's general method. As 
Dirac himself 
emphasized~\cite{dirac}, one of the advantages
 of the canonical approach is that
it 
allows a completely general choice of gauge.
In contrast, gauge
fixed methods tie one to a choice of gauge before it is
apparent whether that gauge is or isn't a good choice. In the
situation of interest here, where the background space-time
is singular, 
the question of 
gauge
choice is especially delicate.  Hence, the canonical
approach is preferable.  

In this Section we provide an overview of the main results.
The technical details are relegated to Appendix 1.  Our starting point is 
the Polyakov action for a $p$-brane described by embedding 
coordinates $x^\mu$ in a 
a background space-time with metric 
$g_{\mu \nu}$:
\be
{\cal S}_p = -{1\over 2} \mu_p \int d^{p+1} \sigma
\sqrt{-\gamma}\left(\gamma^{\alpha \beta}
\partial_\alpha x^\mu \partial_\beta x^\nu g_{\mu \nu} -(p-1) \right),
\labeq{pact}
\ee
where $\mu_p$ is a mass per unit $p$-volume. The $p$-brane worldvolume
has coordinates $\sigma^\alpha$, where $\sigma^0 =\tau$ is the
time and $\sigma^i$, $i=1..p$ are the spatial 
coordinates. 

Variation of the action with respect to $\gamma_{\alpha \beta}$
yields the constraint that for $p\neq 1$,
$\gamma_{\alpha \beta}$ equals
the induced metric $\partial_\alpha x^\mu \partial_\beta x^\nu g_{\mu \nu}$
whereas for $p= 1$ $\gamma_{\alpha \beta}$ is conformal
to the induced metric. Substituting these results
back into the action one obtains the Nambu action for the
embedding coordinates $x^\mu(\sigma^\alpha)$ {\it i.e.,}  $-\mu_p$ times the
induced $p$-brane world volume. We shall go back and forth between
the Polyakov and Nambu forms in this paper.
The former is preferable for quantization but the latter is still
useful for
discussing 
classical solutions.

The simplest case of (\ref{eq:pact}) is $p=0$, 
a $0$-brane or massive particle. Writing
$\gamma_{00}=-e^2$ with $e$ the `einbein', one obtains
\be
{\cal S}_0 = {1\over 2} m\int d \tau
\left(e^{-1}
\dot{x}^\mu \dot{x}^\nu g_{\mu \nu} -e\right),
\labeq{parta}
\ee
where we have set $\mu_0=m$ 
and the dot above a variable indicates a derivative with respect to $\tau$.
Variation with respect to $e$ yields the constraint $e^2= -
\dot{x}^\mu \dot{x}^\nu g_{\mu \nu}$.
The canonical momentum is $p_\mu = m g_{\mu \nu}\dot{x}^\nu e^{-1}$ and
the constraint implies the familiar mass shell
condition $g^{\mu \nu} p_\mu p_\nu = -m^2$. 


The canonical treatment for general $p$
is explained in
Appendix 1. 
The main result is that a $p$-brane obeys $p+1$
constraints, reading
\be
C\equiv \pi_\mu  \pi_\nu g^{\mu \nu}
 +\mu_p^2 {\rm Det}( {x}^\mu_{,i} x^\nu_{,j} g_{\mu \nu})\approx 0, \qquad
C_i\equiv  x^\mu_{,i} \pi_\mu\approx 0,
\labeq{consta}
\ee
where `$\approx 0$' means `weakly zero' in sense of the Dirac canonical
procedure (see Appendix I). Here the brane embedding coordinates
are $x^\mu$ and their conjugate momentum densities are
$\pi_\mu$.
The spatial worldvolume coordinates are 
$\sigma^i$, $i=1,\dots,p$, and the  
corresponding partial derivatives are
denoted $x^\mu_{,i}$.
The quantity
${x}^\mu_{,i} x^\nu_{,j} g_{\mu \nu}$ is 
the induced spatial metric on the $p$-brane. In Appendix 2
we calculate the Poisson bracket algebra of
the constraints (\ref{eq:consta}), showing that
the algebra closes and hence the constraints are all first class.
The constraints (\ref{eq:consta}) are invariant under worldvolume
coordinate transformations.

The Hamiltonian giving the most general 
evolution in worldvolume time $\tau$ is then given by
\be
H=\int d^p \sigma \left({1\over 2} A C  + A^i C_i\right),
\labeq{totalh1}
\ee
with $C$ and $C_i$ given in (\ref{eq:consta}).
 The functions $A$ and $A^{i}$ are
completely arbitrary, reflecting the 
arbitrariness in the choice of worldvolume time and space
coordinates. All coordinate choices related by nonsingular
coordinate transformations give equivalent physical
results. 

For $p=0$, anything 
with a spatial index $i$ can be ignored, except 
the determinant in (\ref{eq:consta}) which is replaced by
unity. The first constraint is then the usual mass shell
condition, and the Hamiltonian is an 
arbitrary function
of $\tau$ times the constraint.
The case of $p=1$, {\it i.e.,}  a string, in Minkowski space-time,
$g_{\mu \nu} = \eta_{\mu \nu}$ is also simple and familiar.
In this case, the 
constraints and the Hamiltonian (\ref{eq:totalh1}) are quadratic. 
The resulting equations of motion are linear and hence
 exactly
solvable. The 
constraints (\ref{eq:consta}) amount
to the usual Virasoro conditions.
In general, the $p+1$ constraints (\ref{eq:consta}) together 
with the $p+1$ free choices
of gauge functions $A$ and $A_i$ reduce
the number of physical coordinates and momenta
to $2(d+1) - 2(p+1)= 2(d-p)$, the correct number of transverse 
degrees of freedom for a $p$-brane in $d$ spatial dimensions.

\section{Winding $p$-Branes in ${\cal M}_C \times R^{d-1}$}

In this paper, we shall study the dynamics of branes which wind
around the compact dimension in ${\cal M}_C \times R^{d-1}$,
the line element for which is given in (\ref{eq:backg}).
This space-time possesses an isometry
$\theta \rightarrow \theta +$constant, so one
can consistently truncate the theory to consider 
$p$-branes which 
wind uniformly around the $\theta$ direction.
Such configurations may be described by 
identifying one of the $p$-brane spatial 
coordinates (the $p$'th spatial coordinate,
 $\sigma^p$ say) with $\theta$ and to simultaneously 
insist that
that $\partial_p x^\mu =\partial_p \pi_\mu=0$. 

Through Hamilton's equations, the constraint
$\theta=\sigma^p$ implies 
that $\pi_\theta=0$. This suggests that we can set 
$\theta=\sigma^p$
and $\pi_\theta=0$ and, hence, dimensionally reduce the $p$-brane
to a $(p-1)$-brane. Detailed confirmation that this
is indeed consistent proceeds as follows.
We compute the Poisson brackets between all the
constraints $C$, $C_i$, $\theta-\sigma^p$ and $\pi_\theta$.
Following the Dirac procedure, we then
attempt 
to build a maximal set of first class constraints.
The constraints $C$ and $C_i$ commute with each other, 
for all $\sigma^i$, but not with 
$\theta-\sigma^p$ and $\pi_\theta$. 
The solution is to remove all
the $\pi_\theta$ and $\theta_{,p}$ terms from $C$ and $C_\theta$
by adding terms involving 
$\pi_\theta$ and $\theta_{,p} -1 = (\theta-\sigma^p)_{,p}$.
The new $C$ and $C_i$ are now
first class since they have weakly vanishing Poisson brackets with all
the constraints, and the remaining second class constraints are
$\theta-\sigma^p$ and $\pi_\theta$. For these,
construction of the Dirac bracket is trivial 
and it amounts simply to canceling the $\theta$ and
$\pi_\theta$ derivatives from the Poisson bracket. 
The conclusion is that we can indeed consistently 
set $\theta=\sigma^p$
and $\pi_\theta=0$.
We shall see in the following section that eliminating
$\theta$ and $\pi_{\theta}$ in this way results directly
in the 
good behavior of the winding modes as $t \rightarrow 0$, 
in contrast with the bad behavior of bulk modes.

The surviving first class constraints for winding $p$-branes are
those obtained by substituting $\theta=\sigma^p$ and 
$\pi_\theta=0$ into the $p$-brane constraints (\ref{eq:consta}),
namely
\be
C\equiv \pi_\mu  \pi_\nu \eta^{\mu \nu}
 +\mu_{p}^2 \theta_0^2 t^2 {\rm Det}( {x}^\mu_{,i} x^\nu_{,j} 
\eta_{\mu \nu})\approx 0;\qquad
C_i\equiv  x^\mu_{,i} \pi_\mu\approx 0,
\labeq{redcon}
\ee
where $i$ and $j$ now run from $1$ to $p-1$ and $\mu$ and
$\nu$ from $0$ to $d$. The $t^2$ term comes from the
$\theta \theta$ component of the ${\cal M}_C \times R^{d-1}$ 
background metric (\ref{eq:backg}).
We have also re-defined the 
momentum density $\pi_\mu$ for the $p-1$-brane to be $\theta_0$
times the momentum density for the $p$-brane 
so that the new Poisson brackets are correctly
normalized to give a $p-1$-dimensional delta function.
The Hamiltonian is again given by the form (\ref{eq:totalh1}) 
with the integral taken over the remaining $p-1$ spatial
coordinates. 

For $p=1$, the 
reduced string is a $d$-dimensional particle.
$\pi_\mu$ is now the momentum $p_\mu$ 
and the determinant appearing 
in (\ref{eq:redcon}) should be interpreted as 
unity. The second constraint is
trivial since there are no remaining spatial
directions. The general Hamiltonian reads:
\be
H_0= A(\tau) \left( p_\mu  p_\nu \eta^{\mu \nu}
 +\mu_{1}^2 \theta_0^2 t^2 \right),
\labeq{hemp}
\ee
where $\mu, \nu$ run from $0$ to $d-1$ and 
$A(\tau)$ is an arbitrary function of $\tau$. We shall
study the quantum field theory for this Hamiltonian in
Section VI.


Comparing 
(\ref{eq:redcon}) with (\ref{eq:consta}), we see that
a $p$-brane which winds around the compact
dimension in ${\cal M}_C \times R^{d-1}$ behaves like
a $p-1$-brane in Minkowski spacetime with a time-dependent
effective
tension $\mu_p \theta_0 |t|$, {\it i.e.,}  the $p$-brane
tension times the size of the compact dimension, $\theta_0 |t|$.

\section{Winding states versus bulk states}
\label{bulk}

We have discussed in detail how in the canonical treatment
the coordinate $\theta$ and conjugate momentum density
$\pi_\theta$
may be eliminated
for $p$-branes winding
uniformly around the compact dimension.
This is physically reasonable, since 
 motion of 
a winding $p$-brane along its own length (i.e. along $\theta$)
is meaningless. This is a crucial 
difference from bulk states. Whereas the metric
on the space of coordinates for bulk states includes 
the $t^2 d\theta^2$ term, the metric on the
space of coordinates for winding states does not. 
As we discuss in detail in Appendix 4, when we quantize
the system the square root of the 
determinant of the metric on the space
of coordinates appears in the quantum field
Hamiltonian. For bulk modes the 
metric on the space of coordinates
inherits the singular behavior of the background
metric 
(\ref{eq:backg}), degenerating at $t=0$ so that 
causing the field equations to 
become singular at $t=0$ even for $\theta$-independent 
field modes (see Section X). Conversely, for winding modes the Hamiltonian
operator is regular at $t=0$. 

The metric on the space of coordinates is defined
by the kinetic energy term in the action: if the action reads
${\cal S} ={1\over 2} \int d\tau g_{IJ}\dot{x}^I \dot{x}^J + \dots$,
where $x^I$ are the coordinates, then $g_{IJ}$ is the metric on
the space of coordinates. The sum over $I$ 
includes integration
over $\sigma$ in our case. 
This superspace metric is needed for quantizing
the theory, for example in the coordinate representation one
needs an inner product on Hilbert space and this 
involves integration over coordinates.
The determinant of $g_{IJ}$ is needed in
order to define this integral (see Appendix 4).

The simplest way to identify the physical degrees of freedom 
is to choose a gauge, for example $A$=constant, $A_i=0$.
For bulk particles, we can then read off the metric 
on coordinate space from the action (\ref{eq:parta}) -
in this case it is simply the background
metric itself. 
 
We have already derived the Hamiltonian for winding states, and
showed how through the use of Dirac brackets 
the $\theta$ coordinate may be discarded. If we choose
the gauge 
$A=1$, $A_i=0$
in the Hamiltonian (\ref{eq:totalh1}) with constraints
given in (\ref{eq:redcon}), we can 
construct the 
corresponding gauge-fixed action:
\be
{\cal S}_{gf}= \int d\tau d^{p-1}\sigma \, \,
{1\over 2} \left(\dot{x}^\mu \dot{x}^\mu \eta_{\mu \nu}
-\mu_{p-1}^2 \theta_0^2 t^2 {\rm Det}( {x}^\mu_{,i} x^\nu_{,j} 
\eta_{\mu \nu})\right)
\labeq{actw}
\ee
where $\mu,\nu $ run over $0$ to $d-1$ and $i,j$ run from
$1$ to $p-1$. One may check that the classical equations 
following from the action (\ref{eq:actw}) are the 
correct Lagrangian equations for the $p$-brane in a certain
worldvolume coordinate system and that these equations preserve the
constraints (\ref{eq:redcon}) (see Appendix 3).

The metric on the space of coordinates may be inferred from
the kinetic term in (\ref{eq:actw}), and it is just
the Minkowski metric. 
In contrast, as discussed, the metric
on the space of coordinates for bulk states involves
the full background metric (\ref{eq:backg}) which
degenerates at $t=0$.
The difference means that whereas the
quantum fields describing winding states are regular in
the neighborhood of $t=0$, those describing bulk
states exhibit
logarithmic
divergences. In the penultimate section of this paper
we argue that these divergences are plausibly the origin of the
bad perturbative behavior displayed by strings
and particles propagating on Lorentzian orbifolds, behavior 
we do not expect to be exhibited in M2-brane winding states
in M theory.

\section{Toy Model: Winding Strings in ${\cal M}_C \times R^{d-1}$}

Before approaching the problem of quantizing winding membranes,
we start with a toy model
consisting of winding string states propagating in
${\cal M}_C \times R^{d-1}$. This problem has also been
considered by others~\cite{bachas,pioline} and
in more detail than we shall
do here. They point out and exploit interesting analogies with
open strings in an electric field. Our focus will
be somewhat different and will serve mainly as a 
warmup for case of winding M2-branes
which we are
more interested in.

Strings winding uniformly
around the compact $\theta$ dimension in (\ref{eq:backg}) appear as
particles from the 
$d-$dimensional point of view. To study the classical
behavior of these particles, it is convenient to 
start from the Nambu action for the string,
\be
{\cal S} = - \mu \int d^2\sigma \sqrt{-{\rm Det}( 
\partial_\alpha x^\mu \partial_\beta x^\nu g_{\mu \nu})},
\labeq{string}
\ee
where $\mu$ is the string tension (to avoid clutter 
we set $\mu_1=\mu$ for
the remainder of this section). The string worldsheet coordinates
are $\sigma^\alpha=(\tau,\sigma)$.

For the winding states we consider, we can set $\theta =\sigma$,
so $0\leq \sigma\leq \theta_0$. We insist that the other space-time
coordinates of the string $x^\mu=(t,\vec{x})$ do not depend on
$\sigma$. It is convenient also to choose the gauge 
$t=\tau$, in which the action (\ref{eq:string}) reduces to
\be
{\cal S} = - \mu \theta_0 \int dt  |t| \sqrt{1-\dot{\vec{x}}\,^2},
\labeq{partnam}
\ee
in which $t$ is now the time, not a coordinate. This is 
the usual square root action for 
a relativistic particle, but with a time-dependent
mass $\mu \theta_0 |t|$. The
canonical momentum is $\vec{p}= \mu \theta_0
|t| \dot{\vec{x}}/\sqrt{1-\dot{\vec{x}}\,^2}$ and the 
classical Hamiltonian generating evolution in the time $t$ is
$H= \sqrt{\vec{p}\,^2+(\mu \theta_0 t)^2}$. This is
regular at $t=0$, indicating that the classical equations
should be regular there.

Due to translation invariance, the canonical momentum
$\vec{p}$
is a constant of the motion. Using this, one obtains the
general solution
\be
\vec{x}= \vec{x_0} + {\vec{p}\over \mu \theta_0} 
{\rm sinh}^{-1}\left(\mu \theta_0 t/|\vec{p}|\right), 
\qquad -\infty <t <\infty,
\labeq{particle}
\ee
according to which the particle moves smoothly through
the singularity. At early and late times the large mass
slows the motion to a crawl.
However, at $t=0$ the particle's mass disappears and 
it instantaneously
reaches the speed of light.
The key point for us is that these winding states
have completely unambiguous evolution across $t=0$, even though
the background metric  (\ref{eq:backg}) is singular there.

Now we turn to quantizing the theory, as a warmup for
the membranes we shall consider in the next section.
The relevant classical Hamiltonian was given 
in (\ref{eq:hemp}): it describes a point particle
with a mass $\mu \theta_0 |t|$. In a general background
space-time, ordering ambiguities appear, which are
reviewed in Appendix 4. However, in the case at hand,
there are no such ambiguities.
The metric on the space of
coordinates is the Minkowski metric $\eta_{\mu \nu}$. 
The standard expression for the momentum operator 
$p_\nu = -i \hbar (\partial /\partial x^\nu)$, and 
the 
Hamiltonian $H$ given in (\ref{eq:hemp}) 
are clearly hermitian under 
integration over coordinate
space, $\int d^d x$. Finally, the background 
curvature $R$ vanishes 
for our background so there
is no curvature term ambiguity either. 

Quantization now proceeds by setting
$p_\mu= - i \partial_\mu$ (we use units in which $\hbar$ is unity)
in the Hamiltonian constraint (\ref{eq:hemp}) which
is now an operator acting on the quantum field $\phi$.
Fourier transforming with respect to 
$\vec{x}$, we obtain
\be
\ddot{\phi} = -\left(\vec{p}\,^2+ (\mu \theta_0 t)^2\right) \phi,
\labeq{KG}
\ee
{\it i.e.,}  the Klein Gordon equation for a particle with a mass
$\mu \theta_0 |t|$. 

Equation (\ref{eq:KG}) is the parabolic cylinder equation. Its
detailed properties are discussed in Ref.~\cite{AS},
whose notation we follow.
We write the time-dependent frequency as $\omega\equiv
\sqrt{\vec{p}\,^2 +(\mu \theta_0 t)^2}$. 
At large times $\mu \theta_0 |t|\gg |\vec{p}\,|$, 
$\omega $ is slowly varying:
$\dot{\omega}/\omega^2 \ll 1$ so all modes follow WKB
evolution. The general solution behaves as a
linear combination of 
$\omega^{-{1\over 2}}$exp$(\pm i \int \omega dt)
\approx 
t^{-{1\over 2}} $exp$\left
(\pm i {1\over 2}\left(\mu \theta_0 t^2
+(\vec{p}\,^2{\rm ln}t)/(\mu \theta_0))\right)\right)$.

For large momentum, $\vec{p}\,^2\gg \mu \theta_0$, 
the WKB approximation
remains valid for all time since $\dot{\omega}/\omega^2$ 
is never large. In the WKB approximation there is no
mode mixing and no particle production.
Therefore for large momentum one expects little 
particle production. Departures from 
WKB are nonperturbative in 
 $\dot{\omega}/\omega^2$, as explicit calculation verifies,
the result scaling as $\sim$
exp$(-|\omega^2/\dot{\omega}|_{max}) 
\sim $ exp$(-\vec{p}\,^2/\mu \theta_0)$, at large $\vec{p}^2$.

The parabolic cylinder functions which behave as positive and
negative frequency modes at large times 
are denoted
$E(a,x)$ and $E^*(a,x)$, 
where $x=\sqrt{2\mu \theta_0} t$ and
$a=-\vec{p}\,^2/(2 \mu \theta_0)$. For
positive $x$ they behave respectively
as $x^{-{1\over 2}}$ exp$(+ i({1\over 4}x^2-a {\rm ln} x)$ and 
$x^{-{1\over 2}}$ exp$(- i({1\over 4}x^2-a {\rm ln} x)$.
Both $E(a,x)$ and $E^*(a,x)$ are analytic at $x=0$. 
They are
uniquely continued to 
negative values through the relation
\be
E(a,-x)= -i e^{\pi a} E(a,x) +i \sqrt{1+e^{2\pi a}} E^*(a,x).
\labeq{cont}
\ee
For $t<0$, $E(a,-\sqrt{2\mu \theta_0}t)$ is the positive frequency
incoming mode. As we extend $t$ to positive values, (\ref{eq:cont})
yields the outgoing solution consisting of a linear combination
of the positive frequency solution $E^*$ and the negative 
frequency solution $E$. The Bogoliubov coefficient~\cite{birrell} $\beta$
for modes of momentum $\vec{p}$ is read off from (\ref{eq:cont}):
\be
\beta=-i e^{\pi a} = -i e^{-{1\over 2} \pi \vec{p}\,^2/(\mu \theta_0)}.
\labeq{beta}
\ee
The result is exponentially suppressed at large
$\vec{p}$, hence, the 
total number of particles per unit volume 
created by passage through the singularity is 
\be
\int {d^{d-1} \vec{p} \over (2\pi)^{d-1}} |\beta|^2 
= \left({\mu \theta_0 \over 2\pi}\right)^{d-1\over 2},
\labeq{ppres}
\ee
which is finite and tends to zero as the rapidity of the 
brane collision is diminished.  

It is interesting to ask 
what happens if we attempt to attach the $t<0$ half of
${\cal M}_C \times R^{d-1}$, (\ref{eq:backg}) 
with rapidity parameter $\theta_0^{in}$, to the upper half of
${\cal M}_C \times R^{d-1}$ with a different rapidity parameter
${\theta}_0^{out}$. After all,
the field equations for general
relativity break down at $t=0$ and, hence, there
is insufficient Cauchy data to uniquely determine the solution
to the future. Hence, it might seem that we have the freedom to
attach a future compactified 
Milne with any parameter ${\theta}_0^{out}$, since this would
still be locally flat away from $t=0$ and,
hence, a legitimate string theory background. However
it is quickly seen that this is not allowed. 
By matching the
field $\phi$ and  its first time derivative $\partial_t \phi$ 
across $t=0$, we can determine the particle production 
in this case.
We find that due to the jump in $\theta_0$, 
the Bogoliubov coefficient $\beta$ behaves like
$((\theta_0^{in})^{1\over 2}-({\theta}_0^{out})^{1\over 2})/
\sqrt{\theta_0^{in} \theta_0^{out}}$, 
at large momentum, independent of $\vec{p}$. 
This implies 
divergent particle production, and indicates that to lowest
order one only obtains sensible results in the analytically-
continued background, which has ${\theta}_0^{out}=\theta_0^{in}$. 
We conclude that to retain 
a physically sensible theory we must have ${\theta}_0^{out}=\theta_0^{in}$, 
at least at lowest order in the interactions. 

\section{Dynamics of Winding M2-branes}

We now turn to the more complicated but far more interesting
case of winding membranes in ${\cal M}_C \times R^{d-1}$.
We have in mind eleven-dimensional M theory,
where the eleventh, M theory dimension shrinks away to 
a point. When this dimension is small but static, 
well known arguments~\cite{witten}
indicate that M theory should tend to a string theory:
type IIA for circle compactification, heterotic
string theory for orbifold compactification.
It is precisely the winding membrane states we are
considering which map onto the string theory states as
the M theory dimension becomes small. What makes this
case specially interesting is that the 
string theory states include the graviton and the dilaton.
Hence, by describing string propagation
across $t=0$ we are describing the propagation
of perturbative gravity across a
singularity, which as explained in Section II is 
an FRW cosmological singularity from the $d$ dimensional
point of view.

A winding 2-brane
is  a string from the $d$-dimensional point of view.
As explained in section III, the Hamiltonian 
for such strings may be expressed in a general gauge
as
\be
H=\int d\sigma \left({A\over 2} \left(\pi_\mu  \pi_\nu \eta^{\mu \nu}
 +\mu_2^2 \theta_0^2 t^2 \eta_{\mu \nu} {x^\mu}' {x^\nu}'\right)
+ A^1 {x^\mu}' \pi_\mu \right),
\labeq{totalh2}
\ee
where $A$ and $A^1$ are arbitrary functions of the
worldsheet coordinates $\sigma$ and $\tau$, and prime 
represents the derivative with respect to $\sigma$.
Here $\mu,\nu$ run over $0,1,..,d-1$, and primes denote derivatives
with respect to $\sigma$.
The Hamiltonian is supplemented by the following first class
constraints
\be
\pi_\mu  \pi_\nu \eta^{\mu \nu}
 +\mu_{2}^2 \theta_0^2 t^2 \eta_{\mu \nu} {x^\mu}' {x^\nu}'
\approx 0;\qquad
 {x^\mu}' \pi_\mu\approx 0,
\labeq{redconr}
\ee
which ensure the Hamiltonian is weakly zero. The latter
constraint is the familiar requirement that the momentum density is normal to
the string.

The key point for us is that the Hamiltonian and the constraints 
are regular in the
neighborhood of
$t=0$, implying that for a generic class of worldsheet coordinates
the solutions of the equations of motion are regular there.

As we did with particles, it is instructive to examine the
classical theory from the point of view of the Nambu
action. We may directly infer the classical action for
winding membranes on ${\cal M}_C \times R^{d-1}$ by 
setting 
$\theta=\sigma^2$ and $\partial_2 t=\partial_2 
\vec{x} =0$. The Nambu action for the 2-brane then
becomes
\be
{\cal S} = - \mu_2\theta_0  
\int d^2\sigma |t| \sqrt{-{\rm Det}(
\partial_\alpha x^\mu \partial_\beta x^\nu \eta_{\mu \nu})},
\labeq{stringmem}
\ee
where $\sigma^\alpha=(\tau,\sigma)$ and $\mu$ runs over $0,..d-1$.
This is precisely the action for a string 
(\ref{eq:string}) in a time-dependent
background $g_{\mu \nu}= \theta_0 |t|\eta_{\mu \nu}$, the appropriate
string-frame background corresponding to ${\cal M}_C \times R^{d-1}$
in M theory~\cite{kosst}.

As a prelude to quantization, let us discuss the classical
evolution of winding membranes across $t=0$.
We can pick
worldsheet coordinates so that $x^0\equiv t=\tau$, and 
$\dot{\vec{x}}\cdot \vec{x}\,'=0$.
In this gauge the Nambu action is:
\be
{\cal S} = - \mu_2 \theta_0 \int dt d\sigma |t| |\vec{x}\,'|
\sqrt{1-\dot{\vec{x}}\,^2},
\labeq{nambu}
\ee 
and the classical equations are
\be
\partial_t \left(\epsilon \dot{\vec{x}}\right)
= \partial_\sigma \left( {t^2\over \epsilon} \partial_\sigma \vec{x}\right),
\qquad \partial_t \epsilon = t {(\vec{x}\,')^2\over \epsilon},
\labeq{stringe}
\ee
where 
\be
\epsilon = |t| \sqrt{\vec{x'}\,^2/(1-\dot{\vec{x}\,^2})}
\labeq{stringf}
\ee
the energy density along the string is $\mu_2 \epsilon$.
It may be checked that the solutions of these
equations are regular at $t=0$, with the 
energy and momentum density being finite there.
The local speed of
the string hits unity at $t=0$, but
there is no ambiguity in the resulting solutions.
The data at $t=0$ consists
of the string coordinates $\vec{x}(\sigma)$ and the 
momentum density $\vec{\pi}(\sigma) = \mu_2 \epsilon \dot{\vec{x}}$,
which must be
normal to $\vec{x}\,'(\sigma)$ but is
otherwise arbitrary. 
This is the same amount of initial data as that
pertaining at any other
time.

Equations (\ref{eq:stringf}) describe
strings in
a $d$ dimensional flat FRW cosmological
background with $ds^2= a(t)^2 \eta_{\mu \nu}$ and
scale factor $a(t) \propto t^{1\over 2}$.
The usual cosmological
intuition is helpful in understanding
the string evolution. The comparison
one must make is between the curvature scale on the
string and the comoving Hubble radius $|t|$.
When $|t|$ is larger than the comoving
curvature scale on the string, the string oscillates as in flat
space-time,  with fixed proper amplitude and frequency.
However,  when $|t|$ falls below the string curvature scale,
the string is `frozen' in comoving coordinates.
This point of view is useful in understanding the
qualitative behavior we shall discuss in the next section.

There is one final point we wish to emphasize. In Section II we
discussed the general Hamiltonian for a $p$-brane
in curved space. As we have just seen, a winding M2-brane
on ${\cal M}_C \times R^{d-1}$ has the same action as a string in 
the background $g_{\mu \nu}= |t|\eta_{\mu \nu}$. We could
have considered this case directly using the methods
of Section II. The constraints (\ref{eq:consta}) for this case
read:
\be
(\theta_0 |t|)^{-1}\eta^{\mu \nu} \pi_\mu \pi_\nu + \mu_2^2 \theta_0 |t| \eta_{\mu 
\nu} 
{x^\mu}' {x^\nu}'\approx 0 ; \qquad {x^\mu}'\pi_\mu\approx 0,
\labeq{strcon}
\ee
and the Hamiltonian would involve an arbitrary linear combination
of the two. These constraints may be compared with those
coming directly from our analysis of 
winding 2-branes, given by (\ref{eq:redcon})
with $p=2$. 
Only the first constraint differs, 
and {\it only by multiplication by
$\theta_0 |t|$}. 
For all nonzero $\theta_0 |t|$, the difference is insignificant:
the constraints are equivalent. 
Multiplication of the Hamiltonian by any
function of the canonical variables merely amounts to 
a re-definition
of worldsheet time. 
This is just as it should be: dimensionally reduced
membranes are strings. However, the membrane viewpoint is
superior in one respect, namely that the background metric
(\ref{eq:backg}) is non-singular in the physical
coordinates (which do not include $\theta$).
That is why 
the membrane Hamiltonian constraint is nonsingular for these states.
The membrane viewpoint tells us we should 
multiply the string Hamiltonian
by $\theta_0|t|$ in order to obtain a string theory which is 
regular at $t=0$. 
Without knowing  about membranes, the naive reaction
might have been to discard the string theory on the basis
that the string frame metric is singular there.

\section{Einstein Gravity versus an  Expansion in  $1/\alpha'$}

We are interested in the behavior of M theory,
considered as a theory of M2-branes, in the
vicinity of $t=0$. The first question is, for what
range of $|t|$ is the string description valid?
The effective string tension $\mu_1$
is given
in terms of the M2-brane tension $\mu_2$ by
\be
\mu_1= \mu_2 \theta_0 |t|.
\labeq{stringt}
\ee
The mass scale of stringy excitations is the string scale $\mu_1^{1\over 2}$.
For the stringy description to be valid, this scale must
be smaller than the mass of Kaluza-Klein excitations,
$(\theta_0 |t|)^{-1}$. This condition reads
\be
|t|<\mu_2^{-{1\over 3}} \theta_0^{-1}.
\labeq{stringtime}
\ee
The second question is: for what range of $|t|$ may
the string theory be approximated by $d$-dimensional
Einstein-dilaton gravity? 
Recall that the string  frame metric is $g_{\mu \nu} = a^2 \eta_{\mu \nu} = |t| 
\eta_{\mu \nu}$,
so the curvature scale is $ \dot{a}/a \sim 1/|t|$.
The approximation holds 
when the curvature scale is smaller than the string scale  $\mu_1^{1\over 2}$, 
which implies
\be
|t|> \mu_2^{-{1\over 3}} \theta_0^{-{1\over 3}}.
\labeq{stringtime2}
\ee
We conclude that for small $\theta_0$ 
there are three relevant regimes.
In units
where the membrane tension $\mu_2$ is unity, they
are as follows. For $|t| > \theta_0^{-1}$ 
the description of $M$ theory in terms of 
eleven dimensional Einstein gravity should hold. 
As $|t|$ falls below $\theta_0^{-1}$, 
the size of the extra dimension falls below
the membrane scale and we go over to the 
ten dimensional string theory description.
In the cosmological scenarios of interest, 
the incoming state is very smooth~\cite{chaos,review},
hence this state should be well-described by 
ten dimensional Einstein-dilaton gravity, which
of course agrees with the eleven dimensional Einstein theory
reduced in the Kaluza-Klein fashion.
However, when $|t|$ falls below $ \theta_0^{-{1\over 3}}$,
the Einstein-dilaton description fails and we must
employ the fundamental description of strings in
order to obtain regular behavior at $t=0$.

Let us consider, then,  the relevant string action. 
As we explained in Section IV, for winding M2-branes
a good gauge choice 
is  $A=1$, $A^1=0$.
The corresponding gauge-fixed
worldsheet action was given in (\ref{eq:actw}): for $p=2$ this reads
\be
{\cal S}_{gf}= \int d\tau d\sigma
{1\over 2}  \eta_{\mu \nu} \left(\dot{x}^\mu \dot{x}^\nu
-\mu_2^2 \theta_0^2 t^2 {x^\mu}' {x^\nu}')\right).
\labeq{newact}
\ee
where
the fields $x^\mu = (t,\vec{x})$ depend on $\sigma$ and $\tau$. 
This action describes a two dimensional field theory
with a quartic interaction. 

We confine ourselves to some preliminary remarks about
the perturbative behavior of (\ref{eq:newact}) 
before proceeding to
non-perturbative calculations in the next section.
In units where $\hbar$ is unity, the
action must be dimensionless. From the $d$-dimensional point of view,
the coordinates 
$x^\mu$ have dimensions of inverse mass and $\mu$ has dimensions
of mass cubed, so $ \sigma/\tau$ must have dimensions of 
mass squared. However, the dimensional analysis relevant to
the quantization of (\ref{eq:newact}) considered as a two dimensional field
theory is quite different: 
the fields $x^\mu$ are dimensionless. From
the quartic term we see that $\mu_2 \theta_0$ is a dimensionless coupling.
suggesting that perturbation theory in
$\mu_2 \theta_0$ should be
renormalizable. The string tension at a time $t_1$ is
$\mu_1 = \mu_2 \theta_0 |t_1|$; the usual $\alpha'$
expansion is then an expansion in the Regge slope parameter $\alpha'
= 1/(2 \pi \mu_1)$, i.e. in negative powers of $\mu_2 \theta_0$. 
Conversely, the perturbation expansion
we are discussing
is an {\it expansion in inverse powers of $\alpha'$}.

These considerations point the way to the resolution of
an apparent conflict between two facts. Winding M2-brane evolution
is as we have seen smooth through $t=0$. We
expect M2-branes to be described by strings near $t=0$.
The low energy approximation to string theory 
is Einstein-dilaton gravity. Yet, 
as noted above, in Einstein-dilaton gravity,
generic perturbations
diverge logarithmically with $|t|$ as $t$ tends to zero. 
The resolution of the paradox
is that general relativity is the leading term in
an expansion in $\alpha'$ for the string theory. 
As we have shown, however, the
 good behavior of string theory is only apparent
in the opposite expansion, in {\it inverse} powers
of $\alpha'$.

In order to evolve 
the incoming
state defined at large $|t|$ where 
general relativity 
is a good 
description, through small $|t|$ where string theory is still
valid but general relativity fails,
we must match the standard $\alpha'$ 
expansion onto the new $1/\alpha'$ expansion we have been
discussing. We defer a detailed discussion of 
this fascinating issue to a future publication.

The action (\ref{eq:newact}) describes a string with a time-dependent
tension, which goes to zero at $t=0$. There is an extensive literature
on the zero-tension limit of string theory (see for example \cite{devega}),
in Minkowski space-time. At zero tension the action 
is (\ref{eq:newact}) but without the second term. This is the action
for an infinite number of massless particles with no interactions.
Quantum mechanically, there is no central charge and no critical
dimension. However, as the tension is introduced, the usual
central charge and critical dimension appear\cite{nicolaidis}.

\section {Worldsheet Instanton Calculation of Loop Production}

The string theory we are discussing, with action (\ref{eq:newact}),
is nonlinear and therefore difficult to solve. We can still
make substantial progress on questions of physical interest by
employing nonperturbative instanton methods. 
One of the most interesting questions 
is whether one can calculate the quantum production of
M2-branes as the universe passes through the big crunch/big bang 
transition. As we shall now show, this is indeed possible 
through Euclidean instanton techniques.

First, let us reproduce the result obtained in our toy model
of winding string production, equation 
(\ref{eq:beta}). 
Equation (\ref{eq:KG}) may be 
re-interpreted
as a time-independent Schrodinger equation, with $t$ being 
the coordinate, describing an
over-the-barrier wave in a upside-down harmonic potential. The 
Bogoliubov 
coefficient is then just the ratio of the 
reflection coefficient $R$ to the transmission coefficient $T$.
For large momentum 
$|\vec{p}\,|$, $R$ is exponentially small and $T$ is close
to unity
since the WKB
approximation holds.
To compute $R$ we employ the following 
approach, described in the
book by Heading~\cite{heading}. (For related
approximation schemes, applied to string production,
see Refs.~\cite{martinec, gubser}.)

\begin{figure}
{\par\centering
\resizebox*{3.in}{3.in}{\includegraphics{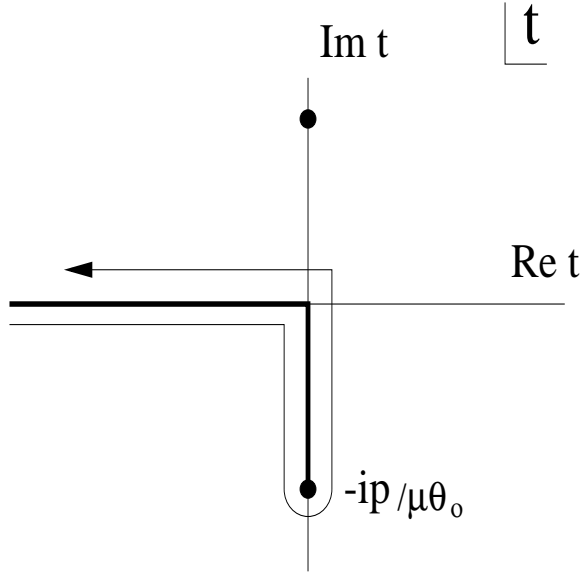}} \vskip .1in \par}
\caption{The contour for computation of Bogoliubov coefficients
in string/membrane production.
}
\labfig{contour}
\end{figure}
\noindent

The method is to analytically
continue the WKB approximate solutions in the complex 
$t$-plane. Defining the WKB frequency, 
$\omega=\sqrt{p^2+(\mu \theta_0 t)^2}$, one
observes there are zeros at $t=\pm i p/(\mu \theta_0)$, where the 
WKB approximation must fail and where the WKB approximate 
solutions possess branch cuts. Heading shows that 
one can compute the reflection coefficient by 
running the branch cut from $t=-ip/(\mu \theta_0)$ 
up the imaginary $t$ axis and out along the negative real $t$ axis. 
Then if one continues an incoming
WKB solution, defined below the cut, in from $t=-\infty$, 
below the 
branch point at $t=-ip/(\mu \theta_0)$ and back out towards 
$t=-\infty$ just above the cut, it becomes the outgoing reflected 
wave. In the leading WKB approximation, the wave is 
given by $w^{-{1\over 2}} e^{-i \int w dt}$. Continuing this
expression along the stated contour, shown in Figure 1,
the exponent acquires a real
contribution on the parts close to the 
imaginary axis, where  $t=-i\tau$. Hence, the magnitude
of the reflection amplitude $R$ is given in the first approximation 
by $e^{(-2 \int_0^p w d \tau)}$ where the factor
of two arises from the two contributions on either side of the
axis.
With $w=\sqrt{p^2-(\mu \theta_0 \tau)^2}$, one easily sees that
the exponent agrees precisely with that in (\ref{eq:beta}).

A more direct method for getting the exponent in the 
Bogoliubov coefficient is to start
not from the field equation (\ref{eq:KG}) but from the 
original action for the fundamental string, (\ref{eq:string}). 
We look for an imaginary-time solution ({\it i.e.,}  an
instanton) corresponding to the WKB continuation described in the
previous section. The 
gauge-fixed particle Hamiltonian was given in (\ref{eq:hemp}), as
$H= {1\over 2} \left(\vec{p}\,^2-p_0^2 +(\mu \theta_0 t)^2\right)$.
The corresponding gauge-fixed action is, in first order form, 
\be
{\cal S} =  \int d \tau (- p_0 \dot{t} + \vec{p} \cdot \dot{\vec{x}} -H),
\labeq{stract}
\ee
where as usual dots over a variable denote $\tau$ derivatives.
Before continuing to imaginary time, it is important
to realize that the spatial momentum $\vec{p}= \dot{\vec{x}}$ 
is conserved (by translation invariance). Hence, all states,
and in particular the asymptotic states we want, are
labeled by $\vec{p}$. We are 
are interested in the transition 
amplitude for fixed initial and final $\vec{p}$,
not $\vec{x}$, and we must use  the appropriate action
which is not (\ref{eq:stract}), but rather
\be
{\cal S} =  \int d \tau (- p_0 \dot{t} - \dot{\vec{p}} \cdot \vec{x} -H),
\labeq{stractm}
\ee
related by an integration by parts. The
$\dot{\vec{p}}\cdot \vec{x}$ term contributes only a phase in the 
Euclidean path integral (because $\vec{p}$ remains 
real) and the $\vec{x}$ integration produces a delta 
function for 
overall momentum conservation. Notice that if we
instead 
had used the naive action $\int d\tau {1\over 2}
\dot{\vec{x}}\,^2$, we would have obtained a $\vec{p}\,^2$ term in
the Euclidean action of
the opposite sign. Similar considerations have been
noted elsewhere~\cite{coleman}.

Now we continue the action (\ref{eq:stractm})
to imaginary time, setting $t=-it_E$
and $\tau=-i\tau_E$. Eliminating $p_0$, the Euclidean action 
${\cal S}_E \equiv -i{\cal S}$) is found to be 
\be
{\cal S}_E= \int d \tau_E \left({1\over 2} 
\left( \dot{t}^2_E -(\mu \theta_0 t_E)^2 +\vec{p}\,^2\right)
+ i \dot{\vec{p}}\cdot\vec{x}\right).
\labeq{eucac}
\ee
where dots now denote derivatives with respect to $\tau_E$. 
The amplitude we want involves $t_E$ running from $0$
to $|\vec{p}\,|$ and back again:  (\ref{eq:eucac}) is just the
action for a simple harmonic oscillator and the required 
instanton 
is $t_E= p \,{\rm cos} (\mu \theta_0 \tau_E)$,
 $-\pi/2 <\mu \theta_0 \tau_E < \pi/2$.
The corresponding Euclidean action is 
\be
{\cal S}_E= {\pi\over 2} {\vec{p}\,^2\over \mu \theta_0},
\labeq{eucacr}
\ee
giving precisely the exponent in (\ref{eq:beta}).

Now we wish to apply this method to calculating the production
of winding membrane states,
described by the action
(\ref{eq:newact}). As in the case of 
analogous calculations of vacuum bubble nucleation 
within field theory \cite{bubbles}, 
it is plausible that objects with the greatest
symmetry are produced since non-symmetrical deformations
will generally yield a larger Euclidean action. Therefore
one might guess that the 
dominant production mechanism is the production of
circular loops. Let us start by considering this case.
The constraint $\pi_\mu {x^\mu}'=0$ implies
that the plane of such loops must be perpendicular to their 
center of mass momentum $p_\mu$. As in the particle production
process previously considered, loops must be produced in
pairs carrying equal and opposite momentum. The Hamiltonian
for such circular loops is straightforwardly found to be 
$H= {1\over 2} \left(\vec{p}\,^2 + 
p_R^2 -p_0^2 +(2 \pi R \mu \theta_0 t)^2\right)$,
Following 
the same steps that led to (\ref{eq:eucac}),
we infer that the appropriate Euclidean action is
\be
{\cal S}_E= \int d \tau_E \left({1\over 2}
\left( \dot{t}^2_E + \dot{R}^2 -(2 \pi \mu_2 \theta_0 R t_E)^2 +\vec{p}\,^2\right)
+ i \dot{\vec{p}}\cdot\vec{x}\right).
\labeq{eucacmem}
\ee
This action describes two degrees of freedom $t_E$ and $R$  interacting
via a positive potential $t_E^2 R^2$. Up to the trivial symmetries
$t_E\rightarrow -t_E$, $R\rightarrow -R$, there is only one
classical solution which satisfying the boundary conditions we
want, namely starting and ending at $t_E=0$, and running up to
the zero of the WKB frequency function at $2 \pi 
\mu \theta_0 R t_E = |\vec{p}\,|$. This solution has $T_E=R$ and 
the Euclidean action is found to be
\be
{\cal S}_E=  {(2 |\vec{p}\,|)^{3\over 2}\over (2 \pi \mu_2 \theta_0)^{1\over 2} }  
\int_0^1 dx \sqrt{1-x^4}.
\labeq{eucacev}
\ee
where the last integral is $
\Gamma[{1\over 4}]^2/(6\sqrt{2 \pi})$, a constant of order unity
which we shall denote $I$.

The Euclidean action 
grows like $|\vec{p}\,|^{3\over 2}$ at large momentum:
this means that the total production of loops is finite. 
Neglecting a possible numerical pre-factor
in the  Bogoliubov coefficient,
we can estimate the number density 
of loops produced per unit volume, 
\be
n  \sim 
\int {d^{d-1} \vec{p} \over (2\pi)^{d-1}} 
e^{-2 S_E} 
= (\mu_2 \theta_0)^{(d-1)/3}  {2^{13-7d\over 3} \Gamma\left(2(d-1)/3\right)
\over 3 \pi^{d-1\over 6}
I^{2(d-1)/3} \Gamma\left((d-1)/2\right)}
\labeq{loops}
\ee
where $I$ is given above.

From the 
instanton solution, the
characteristic size of the loops and the time when they are produced
are both of the same order, 
$R \sim |t| \sim (|\vec{p}\,|/\mu_2 \theta_0)^{1\over 2}\sim
(\mu_2 \theta_0)^{-1/3}$. 
The effective string tension
when they are produced is $\sim \mu_2 \theta_0 |t| 
\sim (\mu_2 \theta_0)^{2\over 3}$. 

We have restricted attention so far to the production of 
circular loops. It is also
important to ask whether long, irregular strings are also 
copiously produced. Even though such strings would be
disfavoured energetically, there is 
an exponentially large density of available states which 
could in principle compensate.
An estimate may be made along the lines of 
Ref.~\cite{gubser}, by simply replacing 
$\vec{p}\,^2$ with $
\vec{p}\,^2 +\mu_1 N$, where $\mu_1$ is the effective
string tension and $N$ is the level number
of the string excitations. 
This picture only makes sense for times greater than 
the string time, so we use the tension at 
the string time, $\mu_1\sim 
(\mu_2 \theta_0)^{2/3}$.
The density of string states
scales as $e^{\sqrt{N}}$ 
hence one should replace (\ref{eq:loops})
with a sum over $N$:
\be
\sim \sum_N e^{\sqrt{N}}
\int d^{d-1} \vec{p} \, 
e^{-\left((\vec{p}\,^2+\mu_1 N)^{3\over 4}/\mu_1^{3\over 4}\right)}.
\labeq{long}
\ee
The $N^{3\over 4}$ beats the $\sqrt{N}$ so the sum is 
dominated by modest $N$, indicating that the production
of long strings is suppressed. According to this result,
the universe emerges at the string time 
with of order one string-scale loop per string-scale
volume, i.e. at a density comparable to but below 
the Hagedorn density.

Another key question is whether
gravitational back-reaction 
effects are likely to be significant 
at the transition. As the universe fills
with string loops, what is their effect on the background geometry?
We estimate this as follows. Consider a string loop
of radius $R$ in M theory frame. Its mass $M$ is $2 \pi R$ times the
effective string tension $\mu_2 L$, where $L$ is the
size of the extra dimension. The effective Einstein-frame 
gravitational coupling (the inverse of the coefficient of 
$R/2$ in the Lagrangian density)
is given 
by $\kappa_{d}^2 = \kappa_{d+1}^2/L$. 
The gravitational potential produced by such a loop 
in $d$ spacetime dimensions is\cite{maeda} 
\be
\Phi = - \kappa_{d}^2{ M\over  ((d-2)A_{d-2} R^{d-3}}
\labeq{gravloop}
\ee
where $A_D$ is the area of the unit $D$-sphere,
$A_D= 2 \pi{D+1\over 2}/\Gamma\left((D+1)/2\right)$.
Specializing to the case of interest, namely 
2-branes in eleven-dimensional M theory, 
the tension $\mu_2$ is related to the eleven dimensional
gravitational coupling by a quantization condition relating
to the four-form flux, reading\cite{duff}
\be
\mu_2^3= 2 \pi^2 /(n \kappa_{11}^2)
\labeq{membten}
\ee
with $n$ an integer. 
Equations (\ref{eq:gravloop}) and (\ref{eq:membten}) 
then imply that the typical gravitational potential around a 
string loop is
\be
\Phi = - {105\over 64 \pi \mu_2^2 R^6 n} 
\sim -(\mu_2^2 R^6 n )^{-1} \sim - \theta_0^2/n
\labeq{deduce}
\ee
up to numerical factors.

We conclude that the gravitational potential on the scale
of the loops 
is of order $\theta_0^2$ and therefore is consistently small
for small collision rapidity.
Since the mean separation of the loops when they are 
produced is of order their size
$R$, this potential $\Phi$ is
the typical gravitational potential throughout space. Multiplying 
the $tt$ component of the background metric (\ref{eq:backg}) by
$1+ 2\Phi$ and redefining $t$, we conclude that the outgoing metric
has an expansion rapidity of 
order $\sim \theta_0(
1+ C \theta_0^2)$ with $C$ a constant of order unity. We conclude that 
for small
$\theta_0$ the gravitational back-reaction due to string loop
production is small. Note that loop production is 
a quantum mechanical effect taking place smoothly over
a time scale of
order $(\mu_2 \theta_0)^{-{1\over 3}}$. Therefore if the
rapidity of the outgoing branes alters as we have estimated,
it happens smoothly and not 
like the jump in $\theta_0$ 
discussed at the end of section VI. Therefore the picture
of loop production is consistent with the comments made there.

\section{Strings on Lorentzian Orbifolds are not regular}

We have shown that strings 
constructed as winding M2-branes
on ${\cal M}_C\times R^{9}$
are analytic in the neighborhood of 
$t=0$. This was the setup originally
envisaged in the ekpyrotic model, where collapse of the
M theory dimension was considered. Subsequently, 
a number of authors investigated the simpler
case of string theory on  ${\cal M}_C\times R^{8}$,
considered as a Lorentzian orbifold solution 
of ten dimensional string theory.
This is a simpler, but 
different setting, hence we expressed 
misgivings\cite{misgive,tolleyperts}
about 
drawing conclusions
from these reduced models.
We shall now explain why the behavior in the
Lorentzian orbifold models
is significantly worse than in M theory
and, hence, why no negative
conclusion should be drawn on the basis of the 
failed perturbative calculations. 

Consider string theory on the background (\ref{eq:backg}). 
Let us choose the gauge $x^0=t=\tau$, and $g_{\mu \nu}
\dot{x}^\mu \dot{x}'^\mu=0$. In this gauge we
express the string solution as $\theta(t,\sigma)$ and $\vec{x}(t,\sigma)$.
If a classical solution in this gauge possesses 
a singularity at $t=0$ in the complex $t$-plane, for all $\sigma$,
then there
can be no choice of worldsheet coordinates $\tau$ and
$\sigma$ which can render the solution analytic in the neighborhood
of $t=0$. For if such 
a choice existed, 
one could re-express $\tau$ in terms of $t$ and, hence, 
$\vec{x}(t,\sigma)$ would be analytic. We shall show that generically,
for strings on Lorentzian orbifolds, the solutions possess
logarithmic singularities {\it i.e.,} 
branch points, rendering them ambiguous as
one circumvents the singularity in the complex $t$-plane.

The demonstration is straightforward, and our argument 
is similar to that in earlier papers\cite{steif}. We are only
interested in the classical equations of motion and
we may compute the relevant Hamiltonian from 
the Nambu action, 
\be
H\equiv \int d\sigma {\cal H} 
= \int d\sigma \sqrt{{\pi_\theta^2\over \mu t^2} + 
{\vec{\pi}^2 \over \mu} + \mu\left( (t {\theta'})^2+{\vec{x}'}\,^2\right)}.
\labeq{loham}
\ee
The Hamiltonian equations allow generic solutions in which 
$\pi_{\theta}(\sigma)$ tends to any 
function of $\sigma$ as $t$ tends to zero. From its
definition, the Hamiltonian density 
${\cal H}$ then diverges as $t^{-1}$. The equation of motion for $\theta$
is
$\dot{\theta} = \pi_\theta/(t^2 {\cal H})$, implying 
that $\dot{\theta} \rightarrow \pm t^{-1}$ independent
of $\sigma$. This implies a leading term
$\theta \sim \pm {\rm log} t$,
independent of $\sigma$. 
Recalling that ${\cal M}_C$ may be rewritten in 
flat coordinates by setting $T=t \cosh \theta$ and $Y=t \sinh \theta$,
one readily understands this behavior. A geodesic in the
$(T,Y)$ coordinates is just $Y= V T$, with $V$ a constant.
At small $t$ this requires $e^\theta$ or $e^{-\theta}$ to
diverge as $t^{-1}$, which is just the result we found.

We conclude that generic solutions to the Hamiltonian
equations possess branch points at $t=0$
meaning that the solutions to the classical
equations are ambiguous as one continues around $t=0$
in the complex $t$-plane. This is a much
worse situation  that encountered for winding
branes in M theory. 

The second problem occurs when we quantize and construct
the associated field theory. Then the bad behavior at $t=0$ 
corresponds to a diverging energy density which renders
perturbation theory invalid.
Since the previous problem 
can be seen even in pointlike states, 
let us focus attention
on those. As we discuss in detail in Appendix 4, one needs
to use the metric on the space of coordinates in order to construct the 
quantum Hamiltonian. In this case, the metric on the space of coordinates
is the background metric, (\ref{eq:backg}). The field equation
for point particles is then given from (\ref{eq:orderh}): Fourier
transforming with respect to $\vec{x}$ and $\theta$, it reads
\be
\ddot{\phi} + {1\over t} \dot{\phi} = -\vec{k}\,^2 \phi - {k_\theta^2\over 
t^{2}} \phi,
\labeq{eoff}
\ee
where $k_\theta$ is quantized in the usual way. This is
the equation studied in earlier work\cite{tolley1} on
quantum field theory on ${\cal M}_C\times R^{d-1}$. The generic
solutions of (\ref{eq:eoff}) behave for small $t$ 
as log~$t$ for $k_\theta=0$,
or $t^{ik_\theta}$ for $k_\theta \neq 0$. In both cases, the
kinetic energy density $\dot{\phi}^2$ 
diverges as $t^{-2}$. Similar behavior is found
for linearized vector and tensor fields on  ${\cal M}_C\times R^{d-1}$.
These divergences lead to the breakdown of perturbation theory
in classical perturbative gravity, 
an effect which is plausibly the
root cause of the 
bad behavior of the associated string theory scattering
amplitudes~\cite{seibergetal}. 
As we have stressed, in the sector of M theory considered as
a theory of membranes, describing perturbative gravity,
this effect does not occur. The field
equations are regular in the neighbourhood of $t=0$ and
there is no associated divergence in the energy density.

With hindsight, one can now see 
that directly constructing string theory on Lorentzian
orbifolds sheds little
light on the M theory case of interest in 
the ekpyrotic and cyclic models. Whilst the orbifolding
construction provides a global map between incoming
and outgoing free fields, it does not
avoid the blueshifting effect which such fields
generically suffer as they approach $t=0$, which
seems to lead to singular behavior in the interactions
(although Ref. \cite{ccosta} argues that a resummation may 
cure this problem).

An alternate approach involving analytic continuation
around $t=0$ has been simultaneously developed, but
so far only implemented successfully in linearized 
cosmological perturbation theory\cite{tolley1,tolleyperts}.
This method may in fact turn out to work even in the M theory
context. The point
is that by circumnavigating
$t=0$ in the complex $t$-plane, maintaining a sufficient
distance from the singularity, one may still retain 
the validity of linear perturbation theory and the
use of the Einstein equations all the way along the
complex time contour. The principles behind this would be
similar
to those familiar in the context of 
WKB matching via analytic continuation.

In any case, the main point we wish to make is that now
that we have what seems like a consistent 
microscopic theory for 
perturbative gravity, valid all the
way through $t=0$, we have a reliable foundation
for such investigations.

\section{Conclusions and Comments}

We have herein proposed an M theoretic model for  the passage through a 
cosmological singularity in terms of a collision of orbifold planes 
in a compactified Milne  ${\cal M}_C\times R^{9}$ background.  The model
begins with two empty, flat, parallel branes a string length apart approaching one 
another at constant rapidity $\theta_0$.  With this initial condition, we have 
argued that the excitations naturally bifurcate into light winding M2-brane modes,  
and a set of massive modes including the Kaluza-Klein massive states.  
It is plausible that the
massive modes decouple as their mass diverges.
The light modes incorporate 
perturbative gravity and, hence, describe the space-time throughout the 
transition.  Our finding that they are produced with finite density following 
dynamical equations that propagate smoothly through the 
transition supports the 
idea that this M-theory picture is 
well-behaved and predictive.  Our model also 
suggests a string theory explanation 
of what goes wrong with
Einstein gravity near the singularity: Einstein gravity is the leading 
approximation in an expansion in $\alpha'$, but the winding 
mode picture is a perturbative expansion in $1/\alpha'$.

Our considerations have been almost entirely classical
or semi-classical, although we believe the
canonical approach we have adopted is a good starting point
for a full quantum theory. 
Much remains to be done to fully
establish the consistency of the picture we are advocating.
In particular, we need to understand the sigma model
in (\ref{eq:actnew}), and make sure that it is 
consistent quantum mechanically in the critical dimension.
Second, we need to
learn how to match the standard $\alpha'$ expansion to the
$1/\alpha'$ expansion which we have argued 
should be smooth around $t=0$. Third, we need to incorporate 
string interactions.   Although the vanishing of
the string theory coupling around $t=0$ suggests that scattering
plays a minor role,
the task of fully constructing string perturbation
theory in this background remains. 

Assuming these nontrivial tasks can be completed, can we 
say something about the significance of this example?
The most obvious application is to the cyclic and  ekpyrotic
universe
models which motivated these investigations,
and which produce precisely the initial conditions required.
As has been argued elsewhere~\cite{kosst,review}, the vicinity
of the collision is well-modeled 
by compactified Milne times flat space. 
Within this model, the calculations reported here 
yield estimates of the reheat temperature immediately
after the brane collision i.e. at the beginning 
of the hot big bang. Furthermore, if
our arguments of Section VIII are correct, they 
should in principle provide 
complete matching rules for evolving cosmological 
perturbations
through singularities of the type
occurring in cyclic/ekpyrotic models. As we discussed
briefly in Section X, it is plausible that this
matching rule will recover the results obtained
earlier for long-wavelength modes 
by the analytic continuation method~\cite{tolley1,tolleyperts},
although that remains to be demonstrated in detail.

Many other questions are raised by our work.
Can we extend the treatment to other time-dependent
singularities?
The background considered here,
${\cal M}_C\times R^{9}$ is certainly very special
being locally flat. Although the string frame metric
is singular, it is conformally flat. 
One would like to
study more generic string backgrounds corresponding to
black holes, or 
Kasner/mixmaster
spacetimes in general relativity. 
As we have argued, there is no reason to
take the latter solutions seriously within M theory since they
are only solutions of the low energy effective theory,
which fails in the relevant regime. Nevertheless, they
presumably have counterparts in M theory, and it remains a 
challenge to find them. We would like to believe that
by constructing one consistent example, namely M theory on
${\cal M}_C\times R^{9}$ we would be opening the door
to an attack on the generic case.

Such a program is admittedly ambitious.  However, should it 
succeed, we believe it
would (and should) completely change our view of cosmology. 
If the best theories of gravity allow for
a smooth passage through time-dependent 
singularities, this must profoundly 
alter our interpretation of the
big bang, and of the major conceptual
problems of the standard hot big bang cosmology.

{\bf Acknowledgments}
We would like to thank L. Boyle,
S. Gratton, S. Gubser, G. Horowitz, S. Kachru, I. Klebanov, 
F. Quevedo, A. Tolley and P. Townsend for helpful remarks.
This work was supported in part by PPARC, UK (NT and MP)
and the US Department of Energy grants DE-FG02-91ER40671 (PJS).
PJS is the Keck Distinguished Visiting
Professor at the Institute for Advanced Study with support from
the Wm.~Keck Foundation and the Monell Foundation and
NT is the Darley Professorial Fellow at Cambridge.

\bigskip

\section*{Appendix 1: Canonical Treatment of $p$-Branes in Curved Space}

In this Appendix we review the canonical treatment of 
$p$-Brane dynamics in curved space. A similar approach is
taken in the recent work of Capovilla {\it et al.}~\cite{capovilla}.
Earlier treatments include Refs.~\cite{townsend} and \cite{gutowski}.

We start from the action in Polyakov form. No square roots appear and the 
ensuing general Hamiltonian involves only polynomial interactions.
The action for a $p$-brane embedded in
a background space-time with coordinates $x^\mu$ and metric
$g_{\mu \nu}$ is:
\be
{\cal S} = -{1\over 2} \mu_p \int d^{p+1} \sigma
\sqrt{-\gamma}\left(\gamma^{\alpha \beta}
\partial_\alpha x^\mu \partial_\beta x^\nu g_{\mu \nu} -(p-1) \right),
\labeq{pacta}
\ee
where $\mu_p$ is a mass per unit $p$-volume. The $p$-brane worldvolume
has coordinates $\sigma^\alpha$, $\alpha=0,1,\dots,p$ in which the
metric is $\gamma_{\alpha \beta}$. 

One way to proceed is to vary (\ref{eq:pacta})
with respect to $\gamma_{\alpha \beta}$, hence, obtaining
the constraint expressing the worldvolume metric 
$\gamma_{\alpha \beta}$ in terms of the 
the induced metric $\partial_\alpha x^\mu \partial_\beta x^\nu g_{\mu \nu}$.
Substituting
back into (\ref{eq:pacta}) one obtains the Nambu action for the
embedding coordinates $x^\mu(\sigma)$ {\it i.e.,}  $-\mu_p$ times the
induced $p$-brane world volume. 

One can do better, however, by not eliminating the worldvolume
metric so soon. Instead it is better to retain
the $\gamma_{\alpha \beta}$ as independent variables and
derive the corresponding 
Hamiltonian and constraints corresponding to evolution
in worldvolume time $\tau$. It is convenient to
express $\gamma_{\alpha \beta}$ in the form frequently
used in canonical general
relativity: the worldsheet line element is 
written
\be
\gamma_{\alpha \beta} d \sigma^\alpha d\sigma^\beta =
(-\alpha^2 +\beta_k \beta^k) d\tau^2 + 2 \beta_i d\tau d\sigma^i 
+\overline{\gamma}_{ij} d\sigma^i d \sigma^j,
\labeq{metric}
\ee
where 
$\sigma^i$,  $i=1,\dots, p$ are the spatial worldvolume coordinates,
$\beta^k$ is the shift vector and
$\alpha$ is the lapse function. The good property of this
representation is that the metric
determinant simplifies: $\gamma=-\alpha^2 \overline{\gamma}$. 

In the following discussion we shall for the most part assume $p>1$ and
then comment on the amendments needed for $p=0,1$.
Using (\ref{eq:metric}) the action (\ref{eq:pacta}) becomes
\be
{\cal S} = -{1\over 2} \mu_p \int d\tau d^{p} \sigma \alpha 
\overline{\gamma}^{1\over 2}
 \left[ - {1\over \alpha^{2}} \dot{x}^\mu  \dot{x}^\nu g_{\mu \nu}
+ 2 {\beta^i \over \alpha^2} \dot{x}^\mu x^\nu_{,i} g_{\mu \nu}
+ (\overline{\gamma}^{ij}-{\beta^i \beta^j \over \alpha^2} )
{x}^\mu_{,i} x^\nu_{,j} g_{\mu \nu} -(p-1)\right].
\labeq{actnew}
\ee
The canonical momenta conjugate to $x^\mu$ are found to be
\be
\pi_\mu = \mu_p {\overline{\gamma}^{1\over 2} \over \alpha} 
(\dot{x}^\nu - \beta^i x^\nu_{,i}) g_{\mu \nu}.
\labeq{momeq}
\ee
No time derivatives of $\alpha$, $\beta^i$ and
$\overline{\gamma}^{ij}$ appear in the Lagrangian, hence, the
corresponding conjugate momenta vanish. In Dirac's language~\cite{dirac},
these are the primary constraints. 
\be
\pi_\alpha\approx 0, \qquad \pi_i\approx 0, \qquad\pi_{ij} \approx 0.
\labeq{firstc}
\ee
Since Poisson brackets between momenta vanish, these constraints 
are first class.
The total Hamiltonian $H$ then consists of the 
usual expression $H\equiv \int d^p \sigma \,[ \pi_A \dot{x}^A -L]$,
\be
H=\int d^p \sigma  \left[{\alpha \over 2 \mu_p \overline{\gamma}^{1\over 2}}
\pi_\mu  \pi_\nu g^{\mu \nu} + \beta^i x^\mu_{,i} \pi_\mu +
{\mu_p \alpha  \overline{\gamma}^{1\over 2} \over 2} 
\left(\overline{\gamma}^{ij}{x}^\mu_{,i} x^\nu_{,j} g_{\mu \nu} -(p-1)\right)
\right],
\labeq{hama}
\ee
plus an arbitrary linear combination of the primary first class
constraints (\ref{eq:firstc}). 

Additional secondary constraints are obtained from the Hamiltonian
equations for $\dot{\pi}_\alpha$, $\dot{\pi}_i$ and $\dot{\pi}_{ij}$.
Insisting these 
vanish as they must for consistency 
with (\ref{eq:firstc}), one finds
\ba
C&\equiv& \pi_\mu  \pi_\nu g^{\mu \nu}
 +\mu_p^2 
\overline{\gamma}
\left(\overline{\gamma}^{ij}{x}^\mu_{,i} x^\nu_{,j} g_{\mu \nu} 
-(p-1)\right)\approx 0,\cr
C_i&\equiv&x^\mu_{,i} \pi_\mu\approx 0,\cr
C_{ij}&\equiv& \gamma_{ij}-{x}^\mu_{,i} x^\nu_{,j} g_{\mu \nu} \approx 0.
\labeq{secondc}
\ea
Following Dirac's procedure, we can now try to eliminate
variables using second class constraints (constraints
whose Poisson brackets with the other constraints
does not vanish). In particular, it
makes sense to eliminate $\gamma_{ij}$
since the corresponding momenta $\pi_{ij}$ vanish weakly.
It is easy to see that $C_{ij}$ are
second class since their Poisson brackets with $\pi_{ij}$ 
are nonzero.
Hence, we eliminate
$\gamma_{ij}$ using the $C_{ij}$ constraint, obtaining
\be
C\equiv \pi_\mu  \pi_\nu g^{\mu \nu}
 +\mu_p^2 {\rm Det}( {x}^\mu_{,i} x^\nu_{,j} g_{\mu \nu})
\approx 0, \qquad 
C_i\equiv x^\mu_{,i} \pi_\mu\approx 0,
\labeq{secondcc}
\ee
as our new constraints, on the remaining variables
$x^\mu$ and $\pi_\mu$. The matrix 
${x}^\mu_{,i} x^\nu_{,j} g_{\mu \nu}$ is the
induced spatial metric on the brane. When written this way, the
$C$ and $C_i$ constraints
have zero Poisson brackets with the remaining 
primary constraints $\pi_\alpha$ and $\pi_i$.
A lengthy but straightforward calculation establishes
that all Poisson brackets between 
$C$ and $C_i$ are weakly vanishing 
(see Appendix 2) and, hence, that  we have a 
complete set of 
first class constraints consisting of $\pi_\alpha$,
$\pi_i$, $C_i$ and $C$.

The canonical Hamiltonian (\ref{eq:hama}) is now seen to be 
a linear combination of $C$ and $C_i$, with coefficients
depending on $\alpha$ and $\beta^i$ respectively. 
The general Hamiltonian consists of a sum of this
term plus an 
arbitrary linear combination of the first class
constraints, 
$\int d \sigma (v_\alpha \pi_\alpha 
+v^i \pi_i)$ where $v_\alpha$ and $v_i$ are arbitrary
functions of the worldvolume coordinates. 
From Hamilton's
equations, one infers that $\dot{\alpha}=v_\alpha$,
and $\dot{\beta}^i=v^i$. Therefore, 
$\alpha$ and $\beta$ are
completely arbitrary functions of time. As Dirac
emphasizes, one can then forget about $\alpha$,
$\pi_\alpha$, $\beta^i$ and $\pi_i$ and just write the
total Hamiltonian for the surviving coordinates as $x^\mu$ and 
$\pi_\mu$ as
\be
H=\int d^p\sigma \left({A\over 2}\left(\pi_\mu  \pi_\nu g^{\mu \nu}
 +\mu_p^2 
{\rm Det}({x}^\mu_{,i} x^\nu_{,j} g_{\mu \nu})\right)+ A^i x^\mu_{,i} \pi_\mu 
\right),
\labeq{totalh}
\ee
i.e. a linear combination of the constraints (\ref{eq:secondcc})
with arbitrary coefficients $A$ and $A^i$.
Different choices of $A$ and  $A^i$ then correspond to different 
choices of worldvolume coordinates.

We count the surviving physical
degrees of
freedom as follows. We start
with the $2(d+1)$ coordinates $x^\mu$ and
momenta $\pi^\mu$, each functions
of the $p+1$ worldvolume coordinates. Then we impose
the $p+1$ constraints
$C=C_i=0$. Finally, in order to specify time evolution 
we must pick $p+1$ 
arbitrary functions 
$A$ and $A_i$.
The remaining physical degrees of
freedom are $2(d+1)-2(p+1)=2(d-p)$ in number, 
the right number of transverse 
coordinates and momenta for the $p$-brane.

We have tacitly assumed $p>1$ in the above analysis. The following
minor amendments are needed for $p=0$ and $1$. For $p=0$,
the worldvolume metric involves $\alpha$ only and one can 
ignore anything with an $i$ index except the determinant,
which is replaced by unity. In particular there
is no integration over $\sigma$ in the Hamiltonian, and
the canonical momentum density $\pi_\mu$ is replaced by the
momentum $p_\mu$.
The only constraint is 
\be
C\equiv  p_\mu  p_\nu g^{\mu \nu}+\mu_0^2\approx 0, 
\labeq{parti}
\ee
which is just the usual mass shell condition. The general
Hamiltonian consists of an arbitrary function of $\tau$ times
$C$.

For $p=1$ the action (\ref{eq:pacta}) is invariant under conformal
transformations of the worldvolume metric, and hence
only two independent
combinations of the three
worldvolume metric variables appear in the decomposition
(\ref{eq:metric}). The corresponding 
two momenta vanish and these are the primary constraints.
Through Hamilton's equations one
finds the 
following secondary constraints: 
\ba
C&\equiv&\pi_\mu  \pi_\nu g^{\mu \nu} +\mu_1^2 {{x}^\mu}' {x^\nu}' g_{\mu 
\nu}\approx 0,\cr
C_1&\equiv&\pi_\mu {{x}^\mu}' \approx 0,
\labeq{strincon}
\ea
where primes denote derivatives with respect to $\sigma^1\equiv \sigma$.
The general Hamiltonian again
takes the form (\ref{eq:totalh}), with $p=1$.

\section*{Appendix 2: Poisson bracket algebra of the constraints}

In the canonical theory \cite{dirac}, one considers arbitrary 
functions of the canonical
variables, and of the time $\tau$.
In our case, the canonical variables are fields
$x^\mu(\sigma)$ and $\pi_\mu(\sigma)$
depending upon $\sigma$, which is regarded
as a continuous index labeling an infinite number
of canonical variables. In particular, the constraints
in (\ref{eq:secondcc}) are infinite in number. In 
this Appendix we shall show that the Poisson bracket
algebra of the constraints closes, and, hence, that, in
Dirac's terminology, they are first class.

The Poisson bracket between any two quantities $M$ and $N$,
which may be arbitrary functions of 
the canonical variables (local or nonlocal in $\sigma$) 
and of the time $\tau$, is
given by
\be
\left\{M,N\right\}\equiv \int d^p\sigma \left(
{\partial M \over \partial x^\mu(\sigma)}
{\partial N\over \partial \pi_\mu(\sigma)}
-{\partial N\over \partial x^\mu(\sigma)}
{\partial M \over \partial \pi_\mu(\sigma)}\right),
\labeq{pb}
\ee
where $\left(\partial x^\mu(\sigma') / \partial x^\nu(\sigma)\right)
= \left(\partial \pi_\nu (\sigma') / \partial \pi_\mu (\sigma)\right)
= \delta^\mu_\nu \delta^p(\sigma-\sigma')$, with other partial
derivatives being zero.

One way to calculate the  Poisson brackets between 
a set of constraints $C$ and $C_i$, 
is to start from a putative Hamiltonian
\be
H=\int d^p \sigma ({A\over 2} C+ A^iC_i),
\labeq{hame}
\ee
where $A$ and $A^i$ are arbitrary functions of $\sigma$, and 
then compute 
Hamilton's equations for the $\tau$ derivatives of
$x^\mu$ and $\pi_\mu$. We then
use these to 
determine the corresponding $\tau$ derivatives of
$C$ and $C_i$. Setting these equal to 
$\{C,H\}$ and 
$\{C_i,H\}$ with $H$ given by (\ref{eq:hame}),
we are able to infer the Poisson brackets between the constraints.
For the Hamiltonian (\ref{eq:hame}) with $C$ and 
$C_i$ given in (\ref{eq:secondcc}), Hamilton's equations read
\ba
\dot{x}^\mu &=& A g^{\mu \nu} \pi_\nu + A^i x^\mu_{\,,i}\cr
\dot{\pi}_\mu &=& (A^i \pi_\mu )_{,i} - {1\over 2} A g^{\lambda \nu}_{\quad,\mu} 
\pi_\lambda
\pi_\nu + \mu_p^2 (A \overline{\gamma}\, \overline{\gamma}
^{ij} x^\nu_{,j})_{,i} g_{\mu \nu}
+ \mu_p^2 A g_{\mu \nu} \Gamma^{\nu}_{\lambda \epsilon} 
 \overline{\gamma}\, \overline{\gamma}^{ij} x^\lambda_{,i}x^\epsilon_{,j},
\labeq{hameqsa}
\ea
where dots denote $\tau$ derivatives and 
$\overline{\gamma}_{ij} ={x}^\mu_{,i} x^\nu_{,j} g_{\mu \nu}$ is the
induced spatial metric on the brane and $ \overline{\gamma}$ its 
determinant. We have made use of the formula
$d  \overline{\gamma}=  \overline{\gamma} \,\overline{\gamma}^{ij} 
d\overline{\gamma}_{ij}$.

Using (\ref{eq:hameqsa}), it is a matter of
straightforward algebra to compute $\dot{C}$ and
$\dot{C_i}$ and hence infer all of 
the Poisson brackets. We find
\ba
\{C(\sigma),C(\sigma')\}&=& 
\left[ (8 \mu_p^2  \overline{\gamma}\, \overline{\gamma}^{ij} 
C_j)(\sigma){\partial
\over \partial \sigma^i} + 4 \mu_p^2 (
 \overline{\gamma}\, \overline{\gamma}^{ij} C_j)_{,i}(\sigma) \right]\delta^p 
(\sigma-\sigma') \cr
\{C(\sigma),C_i(\sigma')\}&=& 
\left[ 2 C(\sigma) {\partial
\over \partial \sigma^i} +C_{,i}(\sigma)\right]
 \delta^p (\sigma-\sigma') \cr
\{C_i(\sigma),C_j(\sigma')\}&=&  
\left[  
C_i(\sigma) {\partial
\over \partial \sigma^j} 
+C_j(\sigma) {\partial
\over \partial \sigma^i} 
+{\partial C_{i}
\over \partial \sigma^j}(\sigma)\right]
\delta^p (\sigma-\sigma'). 
\labeq{results}
\ea
The right hand side consists of linear combinations
of the constraints and, hence, it vanishes weakly.
We conclude that the constraint algebra closes and,
hence that the constraints are first class. Notice that
the case of strings, $p=1$, is specially simple since 
$\overline{\gamma}\, \overline{\gamma}^{11}= 1$ and the
Poisson bracket algebra is linear, with field-independent
structure constants. 

The calculation also provides
a consistency check on our Hamiltonian (\ref{eq:totalh1}),
which is precisely of the form (\ref{eq:hame}), since it implies 
the constraints are preserved under 
time evolution in $\tau$.

\section*{Appendix 3: Equivalence of gauge-fixed Hamiltonian and Lagrangian
equations}

In this Appendix we establish that 
the Lagrangian
equations following from the gauge-fixed action (\ref{eq:newact}) 
for winding $M2$-branes 
are equivalent to the Lagrangian equations for 
a string in the time-dependent background 
$g_{\mu \nu} = |t| \eta_{\mu \nu}$, in a certain string
worldsheet coordinate system.  This is in accord
with our general arguments.

The equations of motion 
following from the gauge-fixed action (\ref{eq:newact})
are:
\ba
\ddot{\vec{x}} &=& t^2 \vec{x}\,''+ 2 t t' \vec{x}\,'\cr
\ddot{t} &=& t \vec{x}\,'\,^2+  t {t'}^2 + t''t^2.
\labeq{new}
\ea
and the constraints take the form
\be
\dot{t} t'=\dot{\vec{x}}\cdot\vec{x}\,'; \qquad \dot{t}^2= \dot{\vec{x}}\,^2 
+ t^2 (\vec{x}'\,^2 -{t'}^2).
\labeq{constrs}
\ee

We want to compare these equations with the Lagrangian
equations of motion  following
from the Polyakov action (\ref{eq:pact}), with $p=1$. These are
\ba
&&\partial_\tau((-\gamma)^{1\over 2}\gamma^{\tau \tau} \partial_\tau x^\mu)
+\partial_\sigma((-\gamma)^{1\over 2} \gamma^{\sigma \sigma} 
\partial_\sigma x^\mu)\cr
&&+(-\gamma)^{1\over 2}
\Gamma_{\nu \lambda}^\mu (\gamma^{\tau \tau}\partial_\tau x^\nu\partial_\tau 
x^\lambda+\gamma^{\sigma \sigma} 
\partial_\sigma x^\nu
\partial_\sigma x^\lambda)=0,
\labeq{strbg}
\ea
where $\Gamma_{\nu \lambda}^\mu$ is the Christoffel symbol for
the background metric. 

We also have the
constraints that the worldsheet metric $\gamma_{\alpha \beta}$
is conformal to the induced metric on the string.
We have the freedom to choose worldsheet coordinates
on the string, but since the equations are conformally
invariant, only the conformal class matters.
The choice $\gamma_{\alpha \beta}
= \Omega^2 {\rm diag}(-t^2, 1)$ is found 
to yield the two constraints (\ref{eq:constrs}).

For our background, $g_{\mu \nu} = |t| \eta_{\mu \nu}$,
we have nonzero Christoffel symbols 
$\Gamma^{0}_{00} = 1/(2t)$, $\Gamma^{i}_{j0} =
\Gamma^{i}_{0j}= \delta_{ij} /(2t)$,
$\Gamma^{0}_{ij} = \delta_{ij}/(2t)$, where $i$ runs over
the background spatial indices $1$ to $d-1$.
The string
equations of motion (\ref{eq:strbg}) are then found
to be equivalent to (\ref{eq:new}), for
all nonzero $t$.

From the string point of view, this choice of gauge 
would seem arbitrary, and indeed it would appear 
to be degenerate at $t=0$. Yet, as we have seen, 
this gauge choice is just $A=1$ and $A^i=0$, which is 
entirely natural from the canonical point of view.
It has the desirable property that
the equations of motion and the constraints are
regular at $t=0$, and from the general properties of
the canonical formalism we are guaranteed the
existence of an infinite class of coordinate
systems, related by nonsingular 
coordinate transformations, 
in which the equations of motion will remain  regular.

\section*{Appendix 4: Ordering ambiguities and their resolution
for relativistic
particles}

In the main text we have discussed the canonical Hamiltonian
treatment of relativistic particles and $p$-branes. When 
one comes to quantize these theories in a general
background, certain ordering ambiguities appear
which must be resolved. Here we provide a brief overview,
following the 
more comprehensive discussion in Ref. \cite{dewitt}.

The field equation for a relativistic particle is simply the
expression of the quantum Hamiltonian constraint $H= 0$, in
a coordinate space representation. The first
task is to determine the representation of the momentum operator
$p_\mu$ in this representation, and then that of the Hamiltonian
operator $H$. As we shall now discuss, this  requires 
knowledge of 
the metric on the space of coordinates. We shall only 
deal with the point particle case. 

The classical Hamiltonian constraint 
for a massive particle
in a background metric $g_{\mu \nu}$ reads
\be
g^{\mu \nu} p_\mu p_\nu + m^2 \approx 0.
\labeq{classpart}
\ee
First we attempt to determine the coordinate space 
representation of  $p_\mu$, consistent
with the quantum bracket:
\be
\left[ x^\mu, p_\nu\right]  =i \hbar \delta_\nu^\mu.
\labeq{braq}
\ee
One choice is $p_\mu = -i\hbar \partial_\mu$
but this is not unique: 
the representation $p_\mu = -i\hbar(\partial_\mu +f_\mu)$,
with $f_\mu$ any function of the coordinates $x^\mu$ and $\tau$,
is equally good as far as (\ref{eq:braq}) is concerned.

We now show how $f_\mu$ may be determined from
the additional requirement that $p_\mu$ be hermitian {\it i.e.,} 
that the momentum be real.
In the coordinate space representation, this requirement reads
\be
\langle \chi| p_\mu |\phi\rangle 
\equiv  \int d^d x (-g(x))^{1\over 2} \left( \chi^* p_\mu \phi\right) 
= \langle \phi| p_\mu |\chi \rangle^* \
\equiv  \int d^d x (-g(x))^{1\over 2} \left( \phi^* p_\mu \chi\right)^*
\labeq{ip}
\ee
where the integration runs over the space of coordinates
and $g_{\mu \nu}$ is the metric on that space.
It is straightforward to check that 
the naive operator  $-i\hbar \partial_\mu$ is in fact {\it not}
hermitian for general $g_{\mu \nu}$, but that 
\be 
p_\mu = -i\hbar \left (\partial_\mu + {1\over 4} \left(\partial_\mu \ln (-
g)\right)\right) = 
-i\hbar (-g)^{-{1\over 4}} \partial_\mu (-g)^{1\over 4}
\labeq{moment}
\ee
is. This discussion uniquely determines the real part of $f_\mu$:
an imaginary part may be absorbed in
an unobservable phase of the wavefunction\cite{dewitt}.

Similarly, when we consider the Hamiltonian constraint (\ref{eq:classpart}),
the questions arise of where to place the $g^{\mu \nu}$ relative
to the $p_\mu$'s, and whether to include
any factors of the metric determinant $g$. The resolution
is familiar: if we write the Hamiltonian in covariant
derivatives on the space of coordinates, it will be hermitian
since we can integrate by parts ignoring the $\sqrt{-g}$ 
factor in the measure. This suggests setting the first term in
(\ref{eq:classpart}) equal to the scalar Laplacian:
\be
g^{\mu \nu} p_\mu p_\nu \rightarrow -\hbar^2 (-g)^{-{1\over 2}}\partial_\mu (-
g)^{1\over 2} g^{\mu \nu} 
\partial_\mu = (-g)^{-{1\over 4}} p_\mu 
(-g)^{1\over 2} g^{\mu \nu} p_\nu  (-g)^{-{1\over 4}}.
\labeq{orderh}
\ee
It is straightforward to check that this is the only
choice of ordering and powers of $(-g)$ which is hermitian and
has the correct classical limit. 
Nevertheless, this 
ordering is not immediately apparent!
More generally, 
one can also include
terms involving commutators of $p_\mu$ which
are zero in the classical limit, but which 
produce the Ricci scalar $R$ in the quantum
Hamiltonian\cite{dewitt}. In the space-time we consider
$R$ is zero. Hence, such
terms do not arise.


\begin{thebibliography}{9999}
\bibitem{kosst}
J.~Khoury, B.A.~Ovrut, N.~Seiberg, P.J.~Steinhardt
and N.~Turok, Phys. Rev. {\bf D65}, 086007 (2002).
\bibitem{steif}
G.T.~Horowitz and A.R.~Steif, Phys. Rev. {\bf D42} (1990) 1950; Phys. Lett. {\bf 
B258} (1991) 91.
\bibitem{tolley1} A.J.~Tolley and N.~Turok, Phys. Rev. {\bf D66} (2002) 106005.

\bibitem{ekperts}
J.~Khoury, B.~A. Ovrut, P.~J. Steinhardt, and N.~Turok,
Phys. Rev. {\bf D66}, 046005 (2002).
\bibitem{greene}
P.~S.~Aspinwall, B.~R.~Greene and D.~R.~Morrison,
hep-th/9309097,
Nucl.\ Phys.\ B {\bf 416}, 414 (1994); E.~Witten, hep-th/9301042,
Nucl.\ Phys.\ B {\bf 403}, 159 (1993).
\bibitem{kost}
J.~Khoury, B.A.~Ovrut, P.J.~Steinhardt, and N.~Turok,
Phys. Rev. {\bf D64}, 123522 (2001).
\bibitem{STu}
 P.J.~Steinhardt and N.~Turok, Science {\bf 296}, 1436 (2002); Phys. 
Rev. {\bf D65}, 126003 (2002).
\bibitem{review}
P.J.~Steinhardt and N.~Turok, astro-ph/0404480.
\bibitem{chaos}
J.K.~Erickson, D.H.~Wesley, P.J.~Steinhardt, and N.~Turok,
Phys. Rev. {\bf D69}, 063514 (2004).
\bibitem{hw}
P.~Ho\v rava and E.~Witten,
Nucl. Phys. {\bf B460} (1996) 506; {\bf B475} (1996) 94.
\bibitem{ovrut} A.~Lukas, B.A.~Ovrut and D.~Waldram,
Nucl. Phys. {\bf B495} (1997) 365.
\bibitem{tolleyperts}
A.J. Tolley, N. Turok and P.~J. Steinhardt, 
Phys. Rev. {\bf D69}, 106005 (2004).
\bibitem{seibergetal} H.~Liu, G.~Moore and N.~Seiberg, JHEP {\bf 0206} (2002) 045;
JHEP {\bf 0210} (2002) 031; O.~Aharony, M.~Fabinger, G.~Horowitz and E.~ 
Silverstein,
hep-th/0204158; M.~Fabinger and J.~McGreevy, hep-th/0206196;
M.~Fabinger and S.~Hellerman, hep-th/0212223; G.T.~Horowitz and J.~Polchinski, 
Phys. Rev. {\bf D66} (2002) 103512; L.~Cornalba and M.S.~Costa, Phys. Rev. {\bf 
D66} (2002) 066001;
L.~Cornalba, M.S.~Costa and C.~Kounnas, Nucl. Phys. {\bf B637} (2002) 378;
L.~Cornalba and M.S.~Costa, hep-th/0302137.
\bibitem{ccosta} 
L.~Cornalba and M.S.~Costa, hep-th/0310099.
\bibitem{dewitt} B.~S. DeWitt, Rev. Mod. Phys. {\bf 29}, 377 (1957);
in {\it Relativity, Groups and Topology},
Les Houches lectures (1963), Gordon and Breach, New York, 1964.
\bibitem{bachas} C.~Bachas, Phys. Lett. {\bf B374} 37 (1996), hep-th/9511043.
\bibitem{pioline}
M.~Berkooz and B.~Pioline, JCAP 0311 (2003) 007, hep-th/0307280; M.~Berkooz,
B.~Pioline and M.~Rozali,
hep-th/0405126.
\bibitem{misgive} N.~Turok, in 
{\it The Future of Theoretical Physics and Cosmology : Celebrating Stephen 
Hawking's 60th Birthday}, eds. G.W.~Gibbons, E.P.S.~Shellard and S.J.~Rankin,
Cambridge University Press, 2003.
\bibitem{dirac} P.A.M. Dirac, {\it Lectures on Quantum Mechanics}, Belfer
Graduate School of Science Monographs Series, New York, 1964.
\bibitem{AS}
M. Abramowitz and I.A. Stegun, {\it Handbook of Mathematical
Functions}, Dover, New York, 1970, page 686.
\bibitem{birrell} See e.g. N.~D. Birrell and P.~C.~W. Davies, {Quantum Fields
in Curved Space-time}, CUP, 1982.
\bibitem{witten} E.~Witten, Nucl. Phys. {\bf B471} (1996) 135;
Nucl. Phys. {\bf B443} (1995) 85.
\bibitem{devega} H.J. de Vega,
I. Giannakis, A.
Nicolaidis, Mod. Phys. Lett. {\bf A10} 2479 (1995); I. Antoniadis and
G. Savvidy, hep-th/0402077 and references therein.
\bibitem{nicolaidis} A. Nicolaidis, J.E. Paschalis, 
P.I. Porfyriadis, 
Phys. Rev. {\bf D58} 047901 (1998);
hep-th/9702185
\bibitem{maeda}
G.W. Gibbons and K. Maeda, Nucl. Phys. {\bf B298} 741 (1988).
\bibitem{duff}
M.J. Duff, TASI Lectures, hep-th/9912164.
\bibitem{heading}
J. Heading, {\it Phase Integral Methods}, Methuen Physical Monographs,
London, 1962, page 85.
\bibitem{martinec}
A.~E. Lawrence and E. Martinec, Class. Quant. Grav. {\bf 13} (1996)
63; hep-th/9509149.
\bibitem{gubser} S.~S. Gubser, hep-th/0305099.
\bibitem{coleman}
K. Lee, Phys. Rev. Lett., {\bf 61} (1988) 263; S. Coleman and
K. Lee, Nuc. Phys. {\bf B329} (1990) 387.
\bibitem{bubbles}
S. Coleman, Phys. Rev. {\bf D15}, 2929 (1977); C. G. Callan Jr.
and S. Coleman, Phys. Rev. {\bf D16}, 1762 (1977).
\bibitem{capovilla} R. Capovilla, J. Guven and E. Rojas, hep-th/0404178;
Nucl. Phys. Proc. Suppl. {\bf 88} 337 (2000).
\bibitem{townsend} J.A.~Azcarraga, J.M.~Izquierdo and P.K.~Townsend,
Phys. Lett. {\bf B267} 366 (1991); Phys. Rev. {\bf D45} 3321 (1992).
\bibitem{gutowski} J.~Gutowski, G.~Papadopoulos and P.K.~Townsend,
Phys.Rev. {\bf D60}  106006 (1999), hep-th/9905156.
\end{thebibliography}
\end{document}